\documentclass[12pt,preprint]{aastex}

\shorttitle{NICMOS Observations of OMC-1}
\shortauthors{Simpson et al.}

\begin{document}

\title{\bf Hubble Space Telescope NICMOS Polarization Measurements of OMC-1}

\author{Janet P. Simpson\altaffilmark{1}, Sean W. J. Colgan, Edwin F. Erickson,}
\affil{NASA Ames Research Center, MS 245-6, Moffett Field, CA 94035-1000} 
\author{Michael G. Burton,} 
\affil{School of Physics, University of New South Wales, Sydney, NSW 2052, Australia}
\author{and A. S. B. Schultz} 
\affil{SETI Institute, 515 N. Whisman Road, Mountain View, CA 94043} 
\email{simpson@cygnus.arc.nasa.gov, Sean.Colgan@nasa.gov, erickson@cygnus.arc.nasa.gov, mgb@phys.unsw.edu.au, schultz@cygnus.arc.nasa.gov}

\altaffiltext{1}{SETI Institute}

\begin{abstract}

We present 2~\micron\ polarization measurements of positions in the BN region 
of the Orion Molecular Cloud (OMC-1) made with 
NICMOS Camera 2 ($0.2''$ resolution) on {\it Hubble Space Telescope}. 
Our goals are to seek the sources of heating for IRc2, 3, 4, and 7, 
identify possible young stellar objects (YSOs), and characterize the grain alignment 
in the dust clouds along the lines-of-sight to the stars.
Our results are as follows: 
BN is $\sim 29$\% polarized by dichroic absorption and appears to be the illuminating source 
for most of the nebulosity to its north and up to $\sim 5''$ to its south.
Although the stars are probably all polarized by dichroic absorption,
there are a number of compact, but non-point-source, objects that could be polarized 
by a combination of both dichroic absorption and local scattering of star light.
We identify several candidate YSOs, including 
an approximately edge-on bipolar YSO $8.7''$ east of BN,
and a deeply-embedded variable star. 
Additional strongly polarized sources are IRc2-B, IRc2-D, and IRc7,
all of which are obviously self-luminous at mid-infrared wavelengths and may be YSOs.
None of these is a reflection nebula illuminated by a star located near radio source I,
as was previously suggested.
Other IRc sources are clearly reflection nebulae:
IRc3 appears to be illuminated by IRc2-B or a combination of the IRc2 sources,
and IRc4 and IRc5 appear to be illuminated by an unseen star in the vicinity of 
radio source I, or by Star n or IRc2-A. 
Trends in the magnetic field direction are inferred from the polarization of the 26 stars 
that are bright enough to be seen as NICMOS point sources. 
Their polarization ranges from $\lesssim 1$\% 
(all stars with this low polarization are optically visible) 
to $> 40$\%.
The most polarized star has a polarization position angle different from its neighbors 
by $\sim 40^\circ$, but in agreement with the grain alignment
inferred from millimeter polarization measurements of the cold dust cloud 
in the southern part of OMC-1.
The polarization position angle of another highly-polarized, probable star 
also requires a grain alignment and magnetic field orientation substantially different 
from the general magnetic field orientation of OMC-1.

\end{abstract}


\section{Introduction}

The Orion Molecular Cloud 1, OMC-1,
centered $1'$ northwest of the Trapezium stars of the Orion Nebula \ion{H}{2} Region,
is the closest and best-studied site of massive star formation.
The brightest near-infrared (NIR) and mid-infrared (MIR) source in the region, 
the Becklin-Neugebauer object BN (Becklin \& Neugebauer 1967),
is possibly a runaway B star from the Trapezium cluster (Plambeck et al.\ 1995; Tan 2004).
It is still surrounded by an optically thick envelope containing silicate dust and ices,
which are identified via their infrared spectral features (e.g., Lee \& Draine 1985; 
Hough et al.\ 1996).
Genzel \& Stutzki (1989) list the indicators of star formation 
in their review of the literature to 1989:
At MIR and far-infrared (FIR) wavelengths, BN is outshone by the extended nebulosity
$\sim 10''$ south called the Kleinmann-Low Nebula (KL, Kleinmann \& Low 1967),
which has a total luminosity of $6 \times 10^4$ to $1.2 \times 10^5 L_\odot$ 
(Genzel \& Stutzki 1989).
In addition to the high luminosity, indicators of massive star formation in the KL region 
include OH, H$_2$O, and CH$_3$OH masers, and 
a dense, warm molecular cloud called the ``hot core'',
which emits an extremely rich spectrum arising from an extensive number of complex molecules
(see Fig.~6 of Genzel \& Stutzki 1989 for a cartoon illustrating the locations of these sources).

At higher spatial resolution, KL is seen to consist of numerous components, all extended,
whose coordinates and finding charts are given by Rieke et al.\ (1973), 
Gezari et al.\ (1998), and Shuping et al.\ (2004).
Controversy has arisen over how many of these ``IRc'' sources, if any, 
contain massive stars that are powering the FIR luminosity.
At L band (3.8 \micron) and MIR wavelengths, the most luminous of these, IRc2, is seen to consist of
at least four compact sources (Chelli et al.\ 1984; Dougados et al.\ 1993; Shuping et al.\ 2004), 
none of which is luminous enough to power the FIR source by itself (Gezari et al.\ 1998).
At NIR wavelengths, the other IRc sources (IRc2 is not readily visible at 2 \micron, 
probably because it is behind the edge of the hot core) are all strongly polarized, 
with centro-symmetric polarization vectors indicating scattered light; that is,
the IRc sources are illuminated by either BN (IRc1) 
or some star in the vicinity of IRc2 (Werner et al.\ 1983;
Minchin et al.\ 1991; Chrysostomou et al.\ 1994).
Most of the fainter IRc sources are probably not self-luminous, but are reflection nebulae 
at NIR wavelengths and warm dust heated by external luminous sources at MIR wavelengths.

On the other hand, radio observations by Menten \& Reid (1996) have identified two
small ionized (possibly \ion{H}{2}) regions: (1) between IRc2 and the hot core 
and (2) at the location of Star ``n'' 
(a very red star first identified in the NIR survey by Lonsdale et al.\ 1982, LBLS).
The star exciting the ionized region, located between IRc2 and the hot core and 
known as radio source ``I'' (Churchwell et al.\ 1987), 
is not detected at any infrared wavelength, 
probably because it is behind too much extinction from the hot core.
Radio source I and Star n are both intriguing because they are  
closer to the centroids of the masers and outflows than the candidate massive stars in IRc2. 
Greenhill et al.\ (2004b) and Menten \& Reid (1995) discuss the SiO masers 
surrounding radio source I,
which they measured with the VLBA and VLA with 0.2 mas and $0.25''$ resolution, respectively.
Greenhill et al.\ (2004b) suggest that source I is excited by a high-mass young stellar object (YSO)
and that the SiO masers arise in the bipolar wind from the surface of its accretion disk.
Star n, about $3''$ southwest of source I, also has a disk seen at MIR wavelengths
(Shuping et al.\ 2004; Greenhill et al.\ 2004a) 
and is near the center of the NH$_3$-line-emitting molecular cloud that includes the hot core 
and another cloud northwest of Star n (Shuping et al.\ 2004).
To summarize, both the star exciting radio source I and Star n appear to have disks --- 
with axes aligned more or less in the northeast-southwest direction --- that may be connected
with the OH and H$_2$O masers in OMC-1.

In addition, there is also a strong outflow, seen in shocked H$_2$, in the northwest-southeast
direction (Allen \& Burton 1993; Schultz et al.\ 1999; Kaifu et al.\ 2000) 
that appears to have its centroid in the IRc2/Star n/source I location. 
It had been speculated that one of these objects is the origin of the outflow; 
however, it is now seen that the outflow does not line up with the outflows
that appear to be associated with sources I and n (e.g., Shuping et al.\ 2004).
We note that the origin of this poorly-collimated H$_2$ outflow is not well-defined 
because of its large extent and wide opening angle, and so deserves further study.
Bally \& Zinnecker (2005) suggest that the outflow is the result of a stellar merger
less than 1000 yr ago; if so, the stars in the region might not currently be ejecting mass
in the northeast and southwest directions.

As discussed above, polarized scattered light has been used to infer 
the spatial locations of sources otherwise invisible due to foreground extinction.
Polarization studies are also useful for tracing the component of the magnetic field 
that lies in the plane of the sky.
Elongated interstellar dust grains in quiescent regions of space 
are aligned with their spin axes (normal to their major axes) parallel to the local magnetic field 
(see e.g., Weingartner \& Draine 2003, Lazarian 2003, and Draine 2004 for reviews 
of the relevant physics).
At FIR to mm wavelengths, the grains emit preferentially along their long axes,
with the result that the emitted light is polarized perpendicular to the local magnetic field.
Houde et al.\ (2004) and Schleuning (1998), at 350 \micron\ and 100 \micron, respectively,
show that the magnetic field in OMC-1 is generally oriented northwest-southeast, 
with the field pinched along the northeast-southwest direction on a scale of several arcmin.
At the location of BN/KL, the polarization percentage is low compared to elsewhere, but that 
may be due to small scale variations undetectable in their $12'' - 35''$ resolution maps.

Light can also be polarized by absorption by aligned grains if the wavelength of the light 
is similar to somewhat larger than the grain size (e.g., Kim \& Martin 1995).
In this case, the elongated grains absorb light 
whose electric vector lies preferentially along their long axes,
so that the transmitted light is polarized parallel to the short axis of the grains.
This is known as dichroic absorption.
Because the short axes of the grains are preferentially aligned with the magnetic field,
the visible and NIR polarization position angles 
are parallel to the component of the magnetic field in the plane of the sky. 
Polarization due to dichroic absorption has long been used to map out 
the orientation of the magnetic field in 
the Milky Way and other galaxies at optical and NIR wavelengths 
(see the review of Milky Way polarization by Heiles \& Crutcher 2005).
The high degree of polarization of BN, proportional to the extinction (Aitken et al.\ 1985, 1989;
 Capps et al.\ 1978; Minchin et al.\ 1991)
and in angular agreement with the FIR-inferred magnetic field directions,
has been used to deduce that BN is polarized by dichroic absorption 
at all NIR and MIR wavelengths where it has been measured 
(e.g., Lee \& Draine 1985; Hough et al.\ 1996; Aitken et al. 1997; Smith et al.\ 2000).

Emission from diffuse nebulosity can also be affected by dichroic absorption.
Hough et al.\ (1986), Burton et al.\ (1991) and Chrysostomou et al.\ (1994) 
measured the polarization of the H$_2$\ $v = 1 - 0$ S(1) line at 2.12 \micron.
They attribute the polarization in the vicinity of IRc2 to dichroic absorption, 
and particularly, the twist in the polarization vectors $5''$ east of IRc2 
to a twist in the magnetic field direction.
On the other hand, they infer that the polarization at large distances from IRc2 and BN 
is due to scattering in an H$_2$ reflection nebula (RN).

In this paper we describe measurements made with 
the Near-Infrared Camera and Multi-Object Spectrometer (NICMOS)
on the {\it Hubble Space Telescope} ({\it HST}) to measure  
the 2~\micron\ polarization of the region at $0.2''$ resolution.
Our region extends $\sim 20''$ north and south of BN; 
in the rest of this paper our references to OMC-1 refer only to this region.
We describe the observations in \S 2, our results on the polarization of both the stars,
and the diffuse nebulosity in \S 3, and summarize our conclusions in \S 4.
We will discuss the illuminating sources of the diffuse nebulosity, several of which are 
probably YSOs, 
and we will discuss the orientation of the magnetic field in OMC-1, as best we can infer it 
from the polarization of the stars.
The locations of the illuminating sources and the magnetic field directions both 
have relevance to the question of the origin of the massive outflows in OMC-1.

\section{Observations}

OMC-1 was imaged with $\it HST$ NICMOS Camera 2 as part of the 
Cycle 12 Program 9752.
The four visits are described in Table 1.
Two visits were of the region north of BN and two south (e.g., the IRc sources);
all four were taken at different telescope orientations so that objects 
otherwise under {\it HST} diffraction spikes could be measured.
Integrations were taken at four dither positions separated by 10.5 pixels ($0.78''$)
in order to remove bad pixels
in each of the three Camera 2 polarizing filters, POL0L, POL120L, and POL240L (hereafter POL),
which cover a 1.9 -- 2.1 \micron\ bandpass.
The detector array was read out in MULTIACCUM mode with sample sequence
STEP16 to accumulate total times of 192 seconds per integration for POL0L and POL120L
and 176 seconds per integration for POL240L, which has better transmission efficiency.
We were able to recover the full dynamic range in this observation
because the design of the STEP16 MULTIACCUM sequence 
incorporates very short dwell times (0.303 s) for the first few read-outs.
Each image was reduced (dark subtracted, flat fielded, etc.) with the HST pipeline (CALNICA).
Because the four-position dither is not adequate to remove certain of the bad pixels,
these pixels (about 50 plus the center column and much of the bottom row)
were removed by substituting the median of surrounding pixels before shifting.
For each polarizing filter, we aligned and shifted the four dither positions
by centroiding two bright stars 
using the IDL program, IDP3\footnote{http://nicmos.as.arizona.edu/software}.
The shifted images were then median-combined to remove the remaining bad pixels; the result is  
that the three images for the three POL filters in each visit 
are all aligned to the same position on the sky.

We used the equations of Hines, Schmidt, \& Schneider (2000) with corrections to the 
filter transmissions for the post-NICMOS Cooling System (installed in 2002 March)
from D. C. Hines (private communication) 
to calculate the Stokes $I$, $Q$, and $U$ intensities. 
Since there are three POL filters, this is done by linear matrix multiplication.
From these we calculated the percentage polarization ($P$), 
polarized intensity ($I_P = P$ times $I$), and position angle $\theta$ for each visit
using the usual relationships 
$P = (Q^2+U^2)^{0.5}/I$ and $\theta = 0.5\ {\rm arctan} (U/Q)$.
The original scale was $0.075948''$ by $0.075355''$ per pixel;
the pixels were rectified and smoothed with a $3\times3$ smoothing function
(i.e., to approximately the {\it HST} spatial resolution of $0.2''$ at 2.0 \micron)
in the data reduction process to achieve higher signal/noise for $P$ and $\theta$.
The intensity images that are plotted in this paper are not smoothed in order to expose 
details such as the fine structure of the NICMOS diffraction pattern,
but the plotted polarization images and the overplotted polarization position angle vectors 
are from the smoothed data.

Since NICMOS launch the flux calibration for the POL filters has been updated only for 
the increased sensitivity resulting from the higher temperatures of 
the NICMOS Cooling System Cryocooler.
Using this calibration, we estimate a noise of 0.02 DN s$^{-1}$ 
or $1.3 \times 10^{-6}$ Jy/pixel in each visit 
from read noise, thermal background,
and source photon noise for the locations of lowest flux.
This corresponds well with the statistical fluctuations in the surface brightness in these regions.
We see no need to correct the data for Cycle 7 NICMOS artifacts such as ``Pedestal'' or
``Residual Shading'', but that may be because the overall high flux levels 
seen everywhere in the images obscure those small effects.

A mosaic of the images from the four visits is shown in Fig.\ 1a.
The mosaic was computed by combining the separate rotated and shifted $I$, $Q$, and $U$ images;
from the combination new values of $P$ and $\theta$ were computed.
This was done for both smoothed and unsmoothed images. 

The brightest object in the whole region is BN;
since it is highly polarized, 
the diffracted polarized flux from it is spread out over many arcsec. 
To remove this diffracted polarized flux, 
we calculated the point-spread function (PSF) for BN using Tiny Tim 
(Krist \& Hook 2004\footnote{http://www.stsci.edu/software/tinytim})
and subtracted the diffracted polarized flux from all the images before further analysis.
This subtraction affects the images at the 0.1\% level at a distance of $\sim 11''$ 
from BN in between the diffraction spikes, and at much higher levels 
in the diffraction spikes, which are not well modeled by Tiny Tim
and which are not included in the mosaics. 
Because we are especially interested in the region around IRc2, we also subtracted the PSFs
computed for Stars n and LBLS ``t''.
These PSFs were computed for 520, 757, and 3400 K black body sources, 
corresponding to the color temperatures of BN, Star n, and Star t, respectively,
which were determined from the H and K magnitudes of Hillenbrand \& Carpenter (2000, HC00)
and Muench et al.\ (2002).

\subsection{Uncertainty Estimates}

During data reduction, CALNICA estimates errors on the intensities per pixel 
measured in each of the POL filters, generally inversely proportional to the photon statistics
with additions for read noise (Mobasher \& Roye 2004).
We assumed that the CALNICA-estimated errors for each exposure 
could be added statistically for median-combining the four dithers 
(i.e., multiplied by $1.18/\sqrt 4$, Snedecor \& Cochran 1980)
to derive the uncertainty per pixel, $\sigma$, for each POL filter.

Sparks \& Axon (1999) discussed the error analysis for a system of three polarizer filters,
such as are found in NICMOS.
They provide equations for the variance and covariance terms of the error matrix for 
$I$, $Q$, and $U$ in terms of the uncertainties $\sigma$ measured in each of the filters.
Sparks \& Axon then give equations for the uncertainties in $\sigma_P$ and $\sigma_\theta$
in terms of these variance and covariance elements.
We used the $\sigma$s for each source measurement in each POL filter  
in Sparks \& Axon's (1999) equations to derive the uncertainties in $P$ and $\theta$.
In general, the uncertainty in $P$ is inversely proportional to $I$,
but the uncertainty in $\theta$ is inversely proportional to $I_P$.

Hines et al.\ (2000) also discussed 
NICMOS error analysis in their instrument paper on NICMOS polarimetry.
They simulated the uncertainties in $\sigma_Q$ and $\sigma_U$ through Monte Carlo calculations 
that included the non-ideal parameters of the NICMOS POL filters
(different transmissions for each filter, imperfect filter orientation),
read noise, photon noise, and the calibration uncertainties.
We interpolated in their curves to estimate $\sigma_Q$ and $\sigma_U$ to use instead of
the CALNICA-calculated $\sigma$s.
For the flux levels found at IRc2 and the faint YSO candidates,
the uncertainties calculated using $\sigma$ values estimated from Hines et al.'s plot
are about twice that of the uncertainties calculated using $\sigma$s from CALNICA.
In this paper we refer to the Hines et al.\ calculated uncertainties as the 
statistical uncertainties.
As an example, for the brightest pixel at the position labeled IRc2-B in Fig.~1, 
with 3 by 3 smoothing for Visit~1
we calculate $I = 5.0$ mJy arcsec$^{-2}$, $P = 0.044$, $\sigma_P = 0.004$, 
$\theta = 121.5^\circ$, and $\sigma_\theta = 2.0^\circ$.

However, in the process of producing the mosaics, 
we noticed that there are some systematic uncertainties in the NICMOS 
POL calibration --- at some locations with certain ranges of $\theta$, the differences 
between the visits ($\theta$ differing by anywhere from 0 to $20^\circ$) 
are much larger than the statistical uncertainties computed  
using the equations of Sparks \& Axon (1999) or Hines et al.\ (2000)
and the estimated errors on the flux computed by CALNICA.
The largest difference between the visits is the area located southwest of IRc3;
differences elsewhere are $< 10^\circ$ (and often much less) 
except for regions affected by the diffraction pattern from BN.
Where the differences between the visits are larger than the statistical errors, 
our quoted uncertainties in $\theta$ are not the computed statistical uncertainties,
but are equal to 0.5 times the differences between the visits.
As a result, except for the faintest stars and faint diffuse sources like IRc2, 
the statistical uncertainties are almost always smaller 
than the differences between the visits.

\section{Results and Discussion}

The stars seen in Fig.\ 1a are identified in Fig.\ 1b.
At all locations the polarization is due to dichroic absorption by aligned dust grains or by 
scattering, usually a combination of both.
North of an east-west line about $5.5''$ south of BN
the dominant appearance of the polarization in the diffuse clouds 
is scattered light from BN,
as seen by the centro-symmetric appearance of the polarization vectors.
At large distances from BN the pattern reverts to or appears to be influenced 
by some dichroic absorption with approximately constant position angle.
Even the ``Crescent'' (Stolovy et al.\ 1998) is illuminated solely by BN.
We will discuss BN and its vicinity further in \S 3.1.2.

South of the line, the region of the KL Nebula, there is no obvious influence 
on the polarized light by BN. From this we conclude
that there is either a line-of-sight component to the distance of BN from KL
or there is a dust lane that blocks the light from BN from reaching this region.
Werner et al. (1983) concluded that there are at least two distinct sources (BN and 
something in the KL Nebula) illuminating the NIR reflection nebulae (RNe); 
however, the sharp distinction between the northern RN illuminated by BN and the southern RNe is 
much more obvious in our data. 

The region south of BN contains the various IRc sources, 
including two components of IRc2: IRc2-B and IRc2-D.
These are shown in Figure 2.
IRc2-B was first detected at a wavelength as short as 2.15 \micron\ by Stolovy et al.\ (1998),
but they did not identify IRc2-D because their data have lower signal/noise owing to
the much smaller bandpass of the NICMOS narrow band filters compared to the POL filters.
Consequently, this is the first detection of any IRc2 component 
at a wavelength as short as 2 \micron, 
and the first published detection of IRc2-D shortward of 3.8 \micron.
These two sub-arcsec sources have source/background ratios of less than 5\%;
a measurement like this is possible only with the high Strehl ratio 
of a space telescope like {\it HST}.
The locations of all four IRc2 sources are marked in Fig.\ 1b,
as are the locations of radio source I and the hot core.
Both IRc2-A and IRc2-C are probably obscured at 2 \micron\ 
by the optically thick hot core that must also obscure the star ionizing source I, 
although all four IRc2 components are visible at 3.8 \micron\ (Dougados et al.\ 1993).

The image in Fig.\ 1a is everywhere polarized to some extent with a fairly constant position angle
except where it is obviously affected by some star's centro-symmetric scattering pattern.
About one third of the continuum at the lowest levels, 
such as in the northeast and southeast sides of the image, 
consists of free-free and bound-free emission from the Orion Nebula \ion{H}{2} region
(estimated from the Pa $\alpha$ image of Schultz et al.\ 1999 and 
the continuum emissivities of Beckert et al.\ 2000).
We will discuss the relative contributions of polarization from scattering and 
dichroic absorption from the foreground material 
in each section as appropriate.

\subsection{Stars and YSOs}

Until this work, the only stars in OMC-1 for which NIR polarization 
had been measured were BN, IRc9 (its diffraction spikes are at $-3''$,$+20''$ in Fig. 1a), 
and Star n (Minchin et al. 1991). 
Thanks to the high spatial resolution and sensitivity of NICMOS,
we have now measured the 2\ micron\ polarization and fluxes 
of a substantial number of stars and candidate YSOs,
including some so faint or confused in the OMC-1 nebulosity that they have no 2 \micron\ 
ground-based measurements. 
In this section we discuss the polarization and fluxes of the most notable objects.
We will use the measured polarization position angles (e.g., Fig.\ 1b) 
to analyse the OMC-1 magnetic field orientation in \S 3.3.

The polarization was measured for each of the stars 
(NICMOS point sources, which are defined as those sources exhibiting
well-defined Airy dark and bright rings from the {\it HST} diffraction pattern);
the results are given in Table 2, along with identifications from optical 
(Hillenbrand 1997, H97), NIR (Lonsdale et al.\ 1982, LBLS; Hillenbrand \& Carpenter 2000, HC00;
Muench et al.\ 2002, MLLA), 
and {\it Chandra} X-ray  (Grosso et al.\ 2005) surveys. 
The star fluxes were measured in each POL filter on the median-combined but 
unshifted and unrotated images.
According to the NICMOS Photometry 
Cookbook\footnote{http://www.stsci.edu/hst/nicmos/performance/photometry},
the calibration for point-source photometry requires that the fluxes be measured in $0.5''$
radius (6.6 pixel) apertures; 
however, the large and variable background in OMC-1 means that the resulting measurements 
are very sensitive to the choice of pixels for background measurements.
As an alternative, the Cookbook suggests measuring the fluxes in much smaller apertures 
and correcting the measurements using a measured or calculated PSF.
This was done with a 2.5 pixel radius for the stellar
measurements and a 5 -- 7 pixel annulus for the background.
These values were chosen because 2.5 pixels is at the minimum of the Airy dark ring at 2 \micron\ 
and the 5 -- 7 pixel annulus is immediately outside the first Airy bright ring.
The aperture correction was estimated by making the same measurements on a PSF computed 
with Tiny Tim (Krist \& Hook 2004) for a 700 K black body source, 
corresponding to the color temperature of highly reddened stars such as these.
At 2 \micron, this correction to the nominal $0.5''$ aperture is a factor of 1.59.

Stokes $I$, $Q$, and $U$ intensities were computed from the fluxes in the three POL filters, and
from these the values of $P$ and $\theta$ in Table 2.
The 2 \micron\ magnitudes in Table 2 were computed from $I$ using a zero magnitude of 734.64 Jy
for the POL filters (STScI Help Desk, private communication)
and the results from the visits were averaged for each star.
The uncertainties in flux, $P$, and $\theta$ were calculated from the uncertainties 
per pixel that were described in \S 2.1 and were propagated in the usual way
for multiple pixels in the apertures and for background subtraction (e.g., Bevington 1967).

The stars detected at visible wavelengths (H97) all show 
low polarization, $\lesssim 2$\%.
The statistical uncertainties in $P$ are of order 1\% or less for all the stars except
the two faintest: Stars  6 and 8, where the uncertainties are 2\%. 
The lowest values of $P$ are probably not real; 
we indicate values of $P$ that probably are real 
by the presence in Table~2 of a value for $\theta$.
Uncertainties in $\theta$ are given in Table~2.
Except for Star 8, which was measured in only one visit, 
the uncertainties in $\theta$ in Table 2 are the differences between the visits, 
since the statistical uncertainties (from Sparks \& Axon 1999; Hines et al.\ 2000; see \S 2.1)
are of the same size or smaller.

Fluxes for the same stars were also obtained from the NICMOS Camera 2 continuum measurements 
at 2.15 \micron, taken 13 April 1997 as an early release observation (PI: E. Erickson); 
the data for these images, published by Stolovy et al.\ (1998), 
were downloaded from the {\it HST} Archive to assure the latest calibration.
These fluxes are also in Table 2.
The fluxes measured in the POL filters at 2 \micron\ have 
an additional zero-point calibration uncertainty 
of 5 -- 10\% over the fluxes measured at 2.15 \micron, 
although the latter data have a much lower signal/noise ratio 
because of the narrow (1\%) F215N filter bandpass compared to the 10\% bandpasses 
of the POL filters.
Most of the differences between the two sets of fluxes is probably due to the effect
of extinction: almost all the stars have larger fluxes at 2.15 \micron\ than at 2.0 \micron. 

To check for variability, 
we compared the magnitudes measured at 2.0 \micron\ with those measured in H and K by HC00
and our measurements of the NICMOS Camera 2 F215N images from 1997.
The only stars that significantly disagree are Star 21 (HC00 755), which is much fainter
in both NICMOS measurements,
and Star 25, which is much brighter and is discussed in \S 3.1.7.
For Star 21, HC00 measured K = 12.436 and H = 12.299 for HC00 755 in February 1999
with an estimated uncertainty of 0.06 mag, 
whereas our M$_{2.0} = 14.96$ refers to data taken in January and August 2004 
and our M$_{2.15} = 14.40$ refers to data taken in April 1997.

Table 2 also contains coordinates of each measured star.
The pixel locations of each star were measured on the rotated $I$ images for each visit 
with respect to the location of BN, which has a core barely larger than the other stars.
With a pixel size of $0.076''$, the position of each star could be measured 
to $\sim 0.01''$ on the image.
Corrections were made for distortion according to the formulas in the NICMOS Data Handbook 
(Mobasher \& Roye 2004).
The dithers were also combined using Drizzle (Koekemoer et al.\ 2002) 
for Visits 1 and 2 (Drizzle could not centroid the images of Visits 3 and 4 because of the 
bright nebulosity) and the star positions were measured on the ``Drizzled'' images.
(The images shown in this paper are not the output from Drizzle because the present images 
have better bad-pixel correction and better registration for the polarization computation.)
We estimate that the uncertainties of the star positions relative to BN are of the order of 
$0.01'' - 0.02''$ from the scatter between the different measurements for two visits 
and the Drizzled images, 
the uncertainties in the NICMOS plate scale, and uncertainties in the distortion correction.
The uncertainties in the absolute positions include the additional uncertainty
in the location of BN, which is given by Rodr\'\i guez et al.\ (2005) to $0.01''$ in 
Right Ascension and Declination for epoch 2002.25.
The coordinates of BN in Table 2 include their measured proper motion,
which agrees well with the proper motion measured for BN by Tan (2004),
to give a position at epoch 2004.65, the date of our Visit 2,
which has the best measurement of BN.
We estimate the final uncertainties in absolute positions  
to be $\sim 0.03''$ for the other stars.

Several objects previously catalogued as stars are seen here to be extended 
(not stars or NICMOS point sources); 
these are listed in Table 3, along with one object in IRc3 that is not extended, 
but does not show the Airy diffraction rings characteristic of NICMOS point sources.
These candidate YSOs are discussed later.

\subsubsection{Star n}

Star n lies at the center of a $< 1''$ double-lobed \ion{H}{2} region  
and is probably responsible for its ionization (Menten \& Reid 1995).
The position angle of the line connecting the two lobes is $\sim 12^\circ$ 
(Menten \& Reid 1995; Greenhill et al.\ 2004a).
A disk approximately perpendicular to the two lobes can be detected at MIR wavelengths 
(Greenhill et al.\ 2004a; Shuping et al.\ 2004); 
NIR spectra show CO in emission, probably originating in the same disk (Luhman et al.\ 2000).
Greenhill et al.\ (2004a) estimate a luminosity of $\sim 2000~L_\odot$, about that of a mid-B star,
perhaps a Herbig AeBe star with a disk.
Feigelson et al.\ (2002) note that the X-ray spectrum may be from two stars 
along the same line of sight, although Grosso et al.\ (2005) suggest
that the X-ray emission comes from a low-mass companion of Star n.
Using adaptive optics on the VLT, Lagrange et al.\ (2004) detect 
a second faint component $0.609''$ distant at PA$ = 256^\circ$ in the J and H bands, 
an alignment significantly different from that of the MIR disk.
A star at this position is under the $\it HST$ diffraction spikes from n 
but it is marginally detected when the PSF is subtracted from n.
This is shown in Figure 2, which is the region from Close Binary 4 (CB4, Stolovy et al.\ 1998) 
to IRc7 with the PSFs from BN and Stars n and t subtracted.

We see no evidence of either the disk or the \ion{H}{2} region in the region immediately surrounding n.
The nebulosity immediately north and west of n in Fig.\ 2 also emits shocked H$_2$, 
as does the the nebulosity northeast of n and north of Star t 
(Lacombe et al.\ 2004; Chrysostomou et al.\ 1997; Stolovy et al.\ 1998).

In the diffuse interstellar medium, the percent polarization is generally 
proportional to the extinction,
with discrepancies from simple proportionality due to variations in the magnetic field orientation
along the line of sight  (Jones 1989; Jones et al.\ 1992).
Following Jones (1989) and Jones et al.\ (1992), we plot in Figure 3 
our measured polarization versus estimates of optical depth at 2.0 \micron, $\tau_{2.0}$, 
for those stars with ground-based measurements in the literature.
Using the extinction cross sections as a function of wavelength from Draine (2003a, 2003b, 2003c),
we estimate $\tau_{2.0} = 1.784 E$(H$-$K) or $\tau_{2.0} = 1.815 E$(K$-$L$'$), 
where $E$(H$-$K) and $E$(K$-$L$'$) are the extinction-caused color excesses.
We estimate these excesses by subtracting 0.3 from H$-$K and 0.4 from K$-$L$'$,
these being the intrinsic colors of middle M red dwarfs (Lada et al.\ 2004).
For Stars 6 and 9, which have no ground-based measurements,
we estimated the H magnitude from the F166N NICMOS Camera 3 image of Schultz et al.\ (1999)
(H equals 18.3 and 17.4, respectively)
and used the 2.15 \micron\ magnitude from Table 2 for K.
Although $\tau_{2.0}$ is probably not underestimated by much, it could be overestimated 
for line-of-sight extinction estimates if there is substantial circumstellar dust
contributing to a star's red color.

Jones (1989) calculated the maximum possible polarization as a function of the ratio, $\eta$, 
of the extinction cross sections perpendicular and parallel to the long axes of elongated grains: 
$P_{max} = {\rm tanh}\ \tau_P$, where $\tau_P = (1 - \eta)\tau/(1+\eta)$
and $\tau$ is the interstellar extinction optical depth.
The solid line in Fig.\ 3 is $P_{max}$ vs. $\tau_{2.0}$ for $\eta = 0.875$,
which produces a line that Jones (1989) found to be a good upper limit to plots
of observed interstellar $P$ vs. $\tau$ at K (2.2 \micron).
The wavelength dependence of $\eta$ is a strong function of the grain size distribution,
but is probably not significant between 2.2 \micron\ and 2.0 \micron.
Other factors besides grain length and width contribute to $\eta$, 
such as efficiency of grain alignment as a function of grain size,
grain size distribution, and magnetic field orientation with respect to the plane of the sky
(Jones et al.\ 1992); 
consequently, it should be regarded as an arbitrary parameter that may be different in 
dense regions like OMC-1 if the average grain shape is different.
However, we see that no stars with $\tau_{2.0} > 1$ have $P$ higher than this line
although a few are close;
thus in this respect, there is no evidence 
that the grain sizes and shapes in OMC-1 are different 
from those of the general interstellar medium.

With the exception of Star n, the stars with $\tau_{2.0} \gtrsim 3$ also have large polarization.
Star n is exceptional, because its polarization is only 2\%, even though its 
measured H$-$K is $\sim 2.7$ and K$-$L$' \sim 3$ (HC00, Muench et al.\ 2002).
Three possibilities come to mind:
(1) Grains in n's disk and in the foreground have relatively little alignment by
the local magnetic field.
If n's high extinction, as inferred from its red color, is due to an approximately edge-on disk,
the grains may not be aligned in the disk.
The motions in the disk are probably turbulent, which would destroy any alignment.
The remnant $P = 2$\% has $\theta = 119^\circ$, approximately the same
as the angle of the overall OMC-1 dichroic absorptive polarization.
(2) Starlight from n towards the earth is intrinsically polarized at some angle offset 
from the general dichroic pattern, for example, perpendicular to the disk seen in the MIR
at position angle $131^\circ$ (Shuping et al.\ 2004).
If this polarized light is then absorbed dichroicly by intervening layers 
at $\sim 120^\circ$, 
the combination would have the effect of decreasing $P$.
Circular polarization (e.g., Martin 1974; Whitney \& Wolff 2002)
would not be produced since the two angles are almost orthogonal. 
In fact, the circular polarization measured at n is very low, $\lesssim 1$\%
(Buscherm\"ohle et al.\ 2005), 
and so this possibility cannot be rejected.
(3) Polarization from dichroic absorption is low over the whole region between IRc2 and n because 
the grain alignment has been disrupted by the outflows seen in H$_2$ and the masers.

\subsubsection{BN}

BN is one of the most polarized of the stars, with $P \sim 29$\%.
Minchin et al.\ (1991) measured $P_{\rm H} =31.2$\% and $P_{\rm K} = 14.2$\% 
with $\theta \sim 114^\circ$ in a $6''$ aperture.
Johnson et al.\ (1981) measured $P_{2.0} = 23$\% and $\theta = 118^\circ$ in 
a $6'' - 8''$ beam.
Interpolation in the measurements of BN's polarization at H and K would give $P_{2.0} \sim 23$\%.
Integrating over a $6''$ aperture, we measure $P = 26$\% and $\theta = 114^\circ$ for Visit 2,
the only visit with BN$~> 3''$ from the edge of the detector array.
The 2.0 \micron\ magnitude in Table~2 is for this visit; the other visits have
somewhat brighter magnitudes but the magnitudes are less reliable because of 
the star's placement near or on the edges of the array.
Magnitudes may be additionally uncertain for BN because the star saturates the NICMOS detectors 
in the first two seconds.

Figure 4 shows the region surrounding BN after the PSF for the BN star itself is subtracted.
There is a line of low flux extending from near BN to the west at position angle $\sim 300^\circ$.
In the NICMOS images of Stolovy et al.\ (1998) and Schultz et al.\ (1999), this line
is seen to extend as far as star LBLS ``a'', $16''$ west and $8''$ north of BN.
We believe it is a foreground dust lane because it is seen in these PSF-subtracted images 
to also extend past BN to the east as a line of low polarization as well as low flux.
Aligned grains in the dust lane, 
which has almost the same position angle as the magnetic field in the region (Fig.\ 1b), 
would polarize the background light by dichroic absorption.
However, the background light, if it is scattered light from BN,
would be itself polarized at an almost perpendicular direction.
These two polarization patterns cancel, leaving the dust lane with low polarization.
The chief reasons for excluding the possibility of the line indicating an edge-on disk
are the large proper motion of BN through the OMC-1 dust clouds 
(Tan 2004; Rodr\'\i guez et al.\ 2005),
which would tend to disrupt the disk,
combined with the large extent of the dust lane ($> 10^4$~AU).
We note that there is a diffuse source of blue-shifted H$_2$ emitting gas 
 at $3.5''$,$-2.0''$ relative to BN and located in front of the dust lane on the east end
(Stolovy et al.\ 1998; Schultz et al.\ 1999;
this is also known as source 143-225, Colgan et al. 2006,
and Clump 8, Chrysostomou et al.\ 1997).

The appearance of the polarization vector position angles around BN in Figs.\ 1 and 4 is not 
completely centro-symmetric. 
To test whether the asymmetry is due to dichroic absorption, 
we removed varying amounts of dichroic absorption 
assuming constant position angle of $\theta = 114^\circ$, the value at BN (Table 2).
To remove dichroic absorption, we approximated the contributions to $Q$ and $U$ due to 
the postulated dichroic absorption as 
$\Delta Q = I P {\rm cos} (2 \theta)$ and
$\Delta U = I P {\rm sin} (2 \theta)$,
and subtracted $\Delta Q$ and $\Delta U$ from $Q$ and $U$, respectively
(for more detailed discussions of radiative transfer of polarized light, 
see Martin 1974, Jones 1989, and Whitney \& Wolff 2002). 
We computed the revised $P$ and $\theta$ from the revised $Q$ and $U$.
Subtracting foreground dichroic absorption with $P \sim 5$\% produces reasonably 
centro-symmetric polarization vector position angles.
At some locations the data are consistent with dichroic $P \sim 10$\%,
while at other locations, especially the region of the Crescent and northwest of the Crescent,
$P \sim 10$\% is too large.
We conclude that the amount of dichroic absorption for the region surrounding BN may be 
somewhat variable, but that it is less than the dichroic absorption at BN itself.
Consequently, the majority of the dichroic absorption seen in BN must be from aligned grains 
local to the star.

We do not detect the probably lower-mass, X-ray emitting star $0.9''$ northwest of BN 
detected by Grosso et al.\ (2005).

\subsubsection{CB4}

Stolovy et al.\ (1998) first noticed that this star (see Fig.\ 2) is a close double, 
probably a binary.
The two stars had consistently different values of $P$ on the two visits
(not evident in Fig.\ 2, in which $P$ is saturated for CB4), which lead us to surmise 
that at least some of the polarization is local to the stars.
On the other hand, the observed value of $\theta$ is $138^\circ$, whereas the 
position angle of the stars with respect to each other is $154^\circ$.
However, through K-band polarization measurements Jensen et al.\ (2004) found 
that the disks of T~Tauri binary systems are moderately but not exactly aligned 
with each other, which means they certainly are not both aligned with their orbital axes.
We do not measure any change in the position angle or separation ($0.35''$)
of the two stars between 1997 and 2004
with $3\sigma$ upper limits to the changes in position angle and separation 
of $1.5^\circ$ and $0.004''$, respectively.
The nebulosity to the east of CB4 has a centro-symmetric polarization vector pattern
with its center at the location of CB4;
from this we can infer that CB4 is at the same distance and is illuminating this nebulosity.

\subsubsection{YSO 147-220 at $8.6''$ E and $2.5''$ N}

Shown in Figure 5, this source, 147-220, is one of the extended, faint objects 
that Stolovy et al.\ (1998) suggested could be protostellar. 
In this paper we identify the non-NICMOS point sources using the Orion Nebula naming 
convention of O'Dell \& Wen (1994).
Fig.\ 5a shows the percentage polarization and Fig.\ 5b the logarithm of $I$.
The actual polarization structure seen in 5a and 5b is not clear because there is so much
foreground dichroic absorption in the region. 
We assumed there is no intrinsic polarization northeast of the YSO 
(where the polarization is a minimum)
and used the observed polarization at that location,  5\% at $\theta = 105^\circ$,
as our estimate of the foreground dichroic absorption contribution.
Using the method for subtracting dichroic absorption described in \S 3.1.2,
we computed revised $P$ and $\theta$.
The revised $P$ is plotted in Fig.\ 5c and the revised polarization vectors 
are plotted on the original $I$ in Fig.\ 5d.
The bipolar nature of this YSO is now clear; 
the star itself is obscured by an optically thick disk 
(it is not a NICMOS point source).

\subsubsection{Candidate YSO 146-231 at $6.8''$ E and  $8.0''$ S}

This small source, 146-231 (see Fig.\ 2), was also first noticed by Stolovy et al.\ (1998).
Because it is so faint (source/background $\sim 0.027$), 
its polarization has large uncertainties ---
it was measured at 100\% and 50\% in Visits 1 and 2, respectively. 
The polarization position angle, $\theta$, is $\sim 88^\circ$, with a statistical uncertainty
of $< 3^\circ$ (calculated as described in \S 2.1).
The uncertainty in Table 3 arises from combining the two visits and is much larger
than this statistical uncertainty.
There are {\it no} sources at the perpendicular angles of $0^\circ$ or $180^\circ$ 
within this uncertainty that could be illuminating this object; 
thus we suggest this object is the diffuse envelope of a low mass YSO.

\subsubsection{Candidate YSO 147-239 at $8.1''$ E and $15.9''$ S}

Figure 6 shows the diffuse source, 147-239,  at $8.1''$E, $15.9''$S 
plus the adjacent Stars 24 and 25 from Table 2.
This object is not obviously protostellar --- it could include scattered light 
from Star 24 in our 2 \micron\ images. 
However, it could instead contain a deeply embedded YSO such that only the light 
forward scattered through the disk is seen. 
This is more likely because it is bright ($L' = 12.615$) at 3.8 \micron,
where it is source 173 of Lada et al.\ (2004).
Unfortunately, Lada et al.\ (2004) did not detect star 24, their source 176, 
because it ``fell into [a] negative `chop' image,''
and thus we cannot compare the colors of the two sources as a function of wavelength.
Whitney \& Wolff (2002) show that an optically thick YSO can appear extended and 
strongly polarized with constant position angle across the source if the grains 
are uniformly aligned within the source.
This may be the case for this object.

\subsubsection{Star 25}

Star 25 is especially interesting (see Fig.\ 6). 
It was about a magnitude and a half brighter in 2004 than in 1997--1998, 
when the images published by Stolovy et al.\ (1998) and Schultz et al.\ (1999) were taken,
and it brightened by 0.3 mag from January, 2004, to August, 2004, when our
two visits were made.
It may be even more variable, if the K magnitude of 14.80 of Muench et al.\ (2002) is accurate and
not ``likely corrupted'', as is given by the note in their table
(because it gives the minimum $\tau_{2.0}$, this is the K value that was used in Fig.\ 3, 
along with L$'$ from Lada et al.\ 2004 to estimate $\tau_{2.0} = 7.8$).
Star 25, with $P = 47$\%, is one of the most polarized stars known at 2 \micron. 
If this polarization is all due to dichroic absorption by grains along the line of sight, 
the grains must be exceptionally well aligned (see Fig.~3)
and/or large compared to other lines of sight, since small grains
are not easily aligned by the interstellar magnetic field 
(e.g., Draine 2003a, 2004; Kim \& Martin 1995).
The polarization position angle of $\sim~2^\circ$ is unusual 
compared to that seen in the other stars;
however, Rao et al.\ (1998) measured the polarization of
a cloud, $5'' - 10''$ in size, at this location containing mm-wave-emitting grains  
with essentially the same grain alignment, 
so Star~25 could be situated behind this cold dust cloud.
Alternatively, Star~25 could have a compact disk ($< 100$ AU, corresponding to $<0.2''$) 
that is polarizing the stellar radiation but shows no extended structure. 
The change in polarization (Table 2) and brightness from Jan. to Aug., 2004, could be due to motion 
of clumps in the disk, increasing the amount of light escaping from the disk.

\subsection{IRc Sources}

The polarization and position angles of the brighter IRc sources 
are given in Table~3.
The source fluxes were measured in each POL filter in software apertures of various radii
(3 to 8 pixels),
and the background was measured in either an annulus surrounding the source 
or in two spots on either side of the source if the annulus included too much extraneous 
nebulosity or diffraction spikes that should not be included in the background.
The annulus method was used for the candidate YSOs and IRc7, 
and the neighboring spot method was used for the IRc2 components.
No background was measured for the locations in IRc3, IRc4, and IRc5 
since they are so bright and extended.
The polarization and position angle were computed with the same matrix transform
as was used for the stellar $P$ and $\theta$ measurements.

\subsubsection{IRc2}

A close-up of the region containing IRc2 is shown in Fig.\ 2.
The perpendiculars to the measured polarization position angles 
for IRc2-B and IRc2-D from Table 3 are $34^\circ \pm 2^\circ$ and $49^\circ \pm 6^\circ$, 
respectively.
Since the position angle from IRc2-B and IRc2-D to radio source I 
(located at $5.99''$,$-7.77''$ with respect to BN,
Menten \& Reid 1995; Rodr\'\i guez et al.\ 2005)
is $179.2^\circ$ and $127.8^\circ$, respectively, we conclude 
that neither IRc2-B or IRc2-D are reflection nebulae illuminated by a star near I.
The position angles from IRc2-B and IRc2-D to Star n are $32.5^\circ$ and $25.4^\circ$,
respectively.
From these angles we see that it is possible that IRc2-B is either a reflection nebula 
illuminated by Star n or
a self-luminous source whose light is polarized by dichroic absorption from foreground 
aligned grains, since the other stars in the vicinity show similar polarization 
position angles and are probably polarized by dichroic absorption.
On the other hand, the polarization position angle of IRc2-D is not consistent 
with either illumination by Star n or with dichroic absorption from grains aligned 
the same as those absorbing IRc2-B, unless the measured $\theta$ is in error by $> 3 \sigma$.
The uncertainty in $\theta$ for IRc2-D is as large as it is 
because of the low source/background ratio ($\sim 0.065$), the low intensity of polarized light, 
and the difficulty in measuring the background.
The argument for the polarization of IRc2-B and -D being due to dichroic absorption 
is strengthened by the measurement of the polarization of IRc2 
at 8 -- 13 \micron\ by Aitken et al.\ (1997) in a $2.6''$ beam.
They separate the components of emissive polarization and absorptive polarization
and find the latter has a position angle of $123^\circ \pm 4^\circ$, in excellent agreement 
with our measurements at 2 \micron.
The situation is much less clear at 3.8 \micron, where Dougados et al.\ (1993),
using a speckle technique, 
measured position angles for the IRc2 components of $\sim 80^\circ$; 
presumably these position angles include some contribution from emissive polarization 
owing to the warm dust content of the objects (Robberto et al.\ 2005).
Considering the brightness and high color temperature seen at MIR wavelengths 
(Gezari et al.\ 1998, 2004; Robberto et al.\ 2005),
we agree with the suggestion that IRc2-B and -D are deeply-embedded, self-luminous sources, 
possibly YSOs, most of whose NIR polarization is produced by dichroic absorption.
We note that IRc2-C was detected in X-rays by Grosso et al.\ (2005), 
consistent with the identification of all the IRc2 sources as candidate YSOs.

\subsubsection{IRc7}

As seen in Figure 2,
the morphology of IRc7 is fan-shaped, with the head pointed towards Star n, 
as though it is caused by an outflow from n (Stolovy et al.\ 1998).
However, the polarization vectors show that IRc7 cannot be illuminated by n in any normal sense 
(perpendiculars to the polarization vectors pointing at the illuminating source).
The polarization of IRc7 could instead be caused by one of the following:

(1) The morphology is indeed due to the outflow from Star n, 
and the polarization is from scattering with n being the illuminating source, but the grains 
are all large ($\gtrsim 0.35$ \micron), such that the scattering angle is $0^\circ$ instead of
the usual $90^\circ$ (this is sometimes called ``polarization reversal'').
Since this size distribution has essentially no small grains ($< 0.35$ \micron), 
we consider it unlikely.

(2) IRc7 is a dust cloud illuminated by a YSO in or near IRc2-B and/or IRc2-D.
IRc7 is certainly not illuminated by any star located near radio source I --- that would
require it to have polarization position angles different by $\sim 17^\circ$ from those measured.
Because of the large polarization of IRc7, 
one could not subtract enough dichroic foreground polarization to change this conclusion
without rotating the low-level surrounding polarization by an unacceptable amount. 

(3) IRc7 contains an embedded self-luminous YSO, the light from which is 
scattered towards us through the aligned grains of the optically-thick IRc7 dust cloud.
The light is then polarized by dichroic absorption with the magnetic field direction 
at position angle $\sim 160^\circ$.
The grains in the IRc7 cloud are aligned with spin axis parallel to the line of sight to n.
The alignment with n may be fortuitous --- one would think that if the alignment is
caused by a wind from n (e.g., Gold 1952),
one would see other effects, such as evidence of shocks,
but IRc7 is one of the few places in OMC-1 where there is no apparent shocked H$_2$
(Stolovy et al.\ 1998; Schultz et al.\ 1999; Lacombe et al.\ 2004).

IRc7 is quite bright at MIR wavelengths, indicating that it contains substantial amounts
of warm dust, with a total luminosity $\sim 400 $ L$_\odot$ (Gezari et al.\ 2004). 
This dust could be heated either by an internal YSO or by absorption of photons from some
external star.
Assuming that all objects are in the plane of the sky 
and the depth of IRc7 is the same as its width, 
we estimate the solid angle (as a fraction of $4\pi$ ster) of IRc7 
as seen from Star n and IRc2 as 0.016 and 0.011, respectively.
If the luminosities of these two objects are
2000 L$_\odot$ (Greenhill et al.\ 2004) and 
1000 -- 5000 L$_\odot$ (Gezari et al.\ 1998, 2004) respectively,
then the most IRc7 can absorb from these sources is
31 L$_\odot$ or 10 -- 50 L$_\odot$,
not nearly enough to power the observed MIR flux of IRc7
unless Star n and the IRc2 sources radiate anisotropically.
Gezari et al.\ (2004) suggest that IRc7 contains an embedded luminous source.
We agree that this seems to be the most likely explanation of our polarization measurements,
that is, the third explanation above.

\subsubsection{IRc3, IRc4, and IRc5}

The bright reflection nebulae, IRc3, IRc4, and IRc5 (Fig.\ 1) have maximum polarizations
of 42\%, 64\%, and 51\%, respectively.
Because of the apparent centro-symmetric position angles of the polarization vectors 
around the IRc2/I region, it has long been thought that one or the other of IRc2 or I
is the illuminating source for these nebulae (e.g., Werner et al.\ 1983; Minchin et al.\ 1991).
The polarization and position angles of a number of parts of these sources 
are given in Table 3.
The fluxes were measured in each POL filter in a 4 to 6 pixel radius aperture,
corresponding to circular apertures of $0.61''$ to $0.91''$,
and $P$ and $\theta$ were computed from the three measured fluxes.
The uncertainties are 0.5 times the differences between the visits,
the standard deviations of the means,  
or $\pm 1$\% for $P$ and $\pm 1^\circ$ for $\theta$, whichever is greater.
We calculated angles from each of these positions to radio source I, IRc2-A, and IRc2-B
to see whether any of these might be the illuminating source for IRc3, 4, and 5.
The illuminating source for the dust cloud IRc3 appears to be IRc2-A, whereas  
a star at the location of radio source I could illuminate IRc4 and IRc5.
This is the only indication of radio source I affecting anything in our images;
however, the difference in the angles from IRc4 or IRc5 to radio source I or IRc2-B
is $<5 - 10^\circ$, or less than 2 to 3 times the uncertainty in the measurement of $\theta$.
Thus, we conclude that the identification of a star at radio source I as the illuminating source 
of IRc4 or IRc5 is not statistically significant.

\subsubsection{Candidate Star, 139-230, in IRc3}

In addition to the one bright star on the southern edge of IRc3, we find a second 
object, 139-230, in the middle of IRc3 that we think is likely to be a star (Table~3).
This object, shown in Figure~7, has a narrow core like the other NICMOS point sources 
but does not have the obvious Airy bright ring 
that we used as our criterion for the NICMOS point sources in Table 2.
The Airy ring, though, would be obscured by the bright nebulosity of IRc3.
If it is a star, its coordinates are ($5^{\rm h} 35^{\rm m} 13\fs90, -5^\circ 22' 29\farcs80$), 
its magnitudes at 2.0 and 2.15 \micron\ are 15.54 and 15.31, respectively, 
and it can be identified as the stellar counterpart of {\it Chandra} source COUP 589 
(Grosso et al.\ 2005).

The uncertainty for $\theta$ for candidate star 139-230, $\pm 6^\circ$, is from the uncertainty
estimate computation described in \S 2.1; 
in this case, the two visits agree to $2^\circ$.
The star's $\theta = 77^\circ \pm 6^\circ$ is greatly different from the $\theta = 122^\circ$ 
of the closest star (Star 18 in Table~2 and Fig.\ 7), 
which has a polarization position angle similar to the other stars of moderate polarization.
The candidate star may be similar to the deeply embedded Star~25 (\S 3.1.7), 
whose polarization position angle agrees with the magnetic field angle of the cold
dust cloud seen at mm wavelengths.
Aitken et al.\ (1997) measured the MIR polarization of the brighter IRc sources in OMC-1.
Because of their 8 -- 22 \micron\ wavelength coverage, they were able to separate 
the absorptive and emissive components of the polarization.
They find $\theta = 87 \pm 5^\circ$ and $44 \pm 2^\circ$ for IRc3 and IRc4, respectively,
for the MIR absorption component, which they attribute to dichroic absorption.
Thus if this candidate star is embedded in or behind IRc3, 
its unusual polarization position angle is not unreasonable.

\subsection{Is the Magnetic Field in OMC-1 Twisted?}

In this section we discuss the direction of the magnetic field in OMC-1, 
making use of our new measurements of the position angles 
determined from dichroic absorption polarization.

On the large scale ($18'' - 35''$ resolution), the magnetic field in Orion 
(determined from FIR polarimetry)
appears regular at position angle $\sim 120^\circ$, but with a slight ``pinch'' 
at a position centered about $45''$ southwest of KL (Schleuning 1998).
The effect of the pinch is to make the field position angle a little larger than $120^\circ$
northwest of KL and a little smaller to the south and east.
We find that the pattern is somewhat different at higher spatial resolution:
according to Fig.\ 1b and Table 2, the average of the more reliable polarization position angles 
to the north and west of Stars n and 18 is $120^\circ \pm 3^\circ$,
whereas the average of the position angles to the southeast (excluding Star 25)
is $139^\circ \pm 2^\circ$.
This shift in position angle could be explained as a possible twist in the magnetic field 
in the plane of the sky at approximately the position of IRc2.

Hough et al.\ (1986), Burton et al.\ (1991), and Chrysostomou et al.\ (1994) 
measured the polarization of the H$_2$ 2.12 \micron\ line, 
which is due to dichroic absorption near the peaks of the line emission 
and to scattered light at great distances from the center.
They find that the polarization position angle changes direction just to the southeast
of IRc2, whereas IRc2 itself is in a region of very low polarization 
(as we also observe, see Fig.\ 2).
Chrysostomou et al.\ suggest that the star producing the northwest-southeast shocked H$_2$ outflow 
has a disk that is threaded poloidally (along the outflow axis) by the OMC-1 magnetic field. 
The field is twisted by the rotation of IRc2's disk, and this warp in the magnetic field
produces both the changes in the polarization position angles southeast of IRc2 and 
the low polarization at IRc2.

Our high spatial resolution observations show that the polarization vectors are sharply twisted
in the region east of IRc2.
However, the 2~\micron\ polarization probably includes additional contribution from scattered light,
most likely from CB4.
Because of this contribution from scattered light, 
we cannot infer any extra twist in the magnetic field in this region.

Plambeck et al.\ (2003) measured the polarization position angles 
of the SiO $v = 0$ masers at radio source I and corrected them for Faraday rotation,
giving a position angle of $\sim~55^\circ$. 
For SiO masers, the polarization position angle can be 
either parallel or perpendicular to the magnetic field (Goldreich et al.\ 1973).
Plambeck et al.\ suggest that polarization is perpendicular to the field, 
such that the field position angle is $\sim 145^\circ$
and the field is parallel to the axis of a disk seen around radio source I.
This model with the disk axis and outflow in the northwest-southeast directions
does not agree with the latest model of source I by Greenhill et al.\ (2004b),
which has the outflow, as determined from the velocities of the SiO and H$_2$O masers,
in the northeast-southwest direction.
However, a field direction of $\sim 145^\circ$ is in good agreement with 
the field direction inferred from
the polarization position angles that we measure in this region.

On the other hand, because we can measure the polarization of very faint stars,
we can measure the polarization of stars at much higher optical depths into the cloud
than has previously been done from the ground.
Figure 8 shows the polarization position angles of those sources 
for which we estimate $\tau_{2.0} > 4$ (\S 3.1.1),
plus the 10~\micron\ absorptive polarization position angles of IRc3 and IRc4 
from Aitken et al.\ (1997)
and the magnetic field position angle at radio source I from Plambeck et al.\ (2003).
The ``anomalous polarization'' region of Rao et al.\ (1998) has essentially the same 
aligned-grain spin-axis direction (close to north-south) as our Star 25 and extends from 
the hot core south by $\sim 10''$.
Figure 2 of Rao et al.\ also has some vertical polarization vectors at approximately the location
of IRc3, from which one could infer approximately horizontal absorptive position angles
in agreement with the NIR (candidate star 139-230 from Table 3) 
and MIR (Aitken et al.\ 1997) measurements.

Do these anomalous polarization regions indicate that the outflow from source I
is affecting the magnetic field in this  region?
The position angles at IRc3, IRc4, and Star 25 in Fig.\ 7 are parallel to extensions of 
the northeast-southwest outflow from source I modeled by Greenhill et al.\ (2004b).
The low polarization at Star n is compatible with this if the line of sight to n 
actually contains two components with almost perpendicular (and cancelling) grain alignments.

Rao et al.\ (1998) suggested that the grains in the region south of the hot core are aligned by 
the high-velocity bipolar outflow seen in CO (e.g., Rodr\'\i guez-Franco et al.\ 1999);
the orientation of this flow is roughly northwest-southeast, similar to the orientation of  
the H$_2$ fingers and with similar orientation to the $120^\circ - 140^\circ$ direction
of the overall magnetic field as determined from polarization measurements.
If the grains are aligned by the recent outflow,
the long axes of the grains would be parallel to the magnetic field instead of perpendicular,
causing the directions of the polarization vectors in this region to be different by $90^\circ$
from elsewhere in OMC-1.
Such a case could happen 
if the grains are aligned by gas streaming past them (long axis parallel to the gas flow, Gold 1952) 
and the gas is streaming at an angle $\lesssim 55^\circ$ to the magnetic field lines 
(with corrections for grain shape, etc., Lazarian 1997).
This short-lived case could occur 
because the timescales for alignment of the largest moment of inertia with the spin axis 
and for the precession of the spin axis around the magnetic field direction are both very short. 
The timescale for the alignment of the spin axis with the magnetic field is larger, and
thus over time, a grain that started out with its spin axis perpendicular to the field direction 
realigns its spin axis parallel to the field direction, 
producing normal grain alignment (spin parallel to magnetic field direction)
(Davis \& Greenstein 1951; Draine 2004 and Lazarian 2003 and references therein 
review other requirements for efficient alignment).

Rao et al.'s (1998) position angle measurement uncertainties are sufficiently large that 
their inferred grain orientations could be in agreement with the overall magnetic field direction
if the grains are aligned as they suggest.
However, our uncertainties (and those of Aitken et al.\ 1997) are sufficiently smaller 
that the position angles of the magnetic field inferred from the polarization 
of 139-230 in IRc3 and Star 25 cannot be brought into agreement 
with this overall magnetic field direction.
We propose that at great depths into the cloud, the magnetic field direction
is twisted with respect to the direction observed in the dichroic polarization of foreground stars,
possibly from effects of the outflow centered in the vicinity of source I or IRc2.
The magnetic field direction deep in the cloud 
would be parallel to the outflow direction 
in both the dust cloud south of the hot core and in IRc3.
The magnetic field direction would not need to be different in IRc4, although the grains
there would have their long axes parallel to the field instead of perpendicular.

\section{Conclusions}

This paper presents 2 \micron\ polarization measurements of OMC-1 
made with NICMOS on {\it HST}.
Thanks to the unprecedented spatial resolution and sensitivity over the whole field of view,
we are able to measure the polarization of 25 stars and 6 YSOs and candidate YSOs 
not previously measured at 2 \micron.
Individual objects discussed in this paper are summarized in Table~4.

The field of view can be divided into two sections by an approximately east-west 
line $\sim 5''$ south of BN.  
BN illuminates the northern region 
but has no apparent influence on the polarization of the scattered light south of $-5.5''$.
From this we conclude that BN is separated from the southern region, 
possibly by a substantial distance in the line of sight.
Another possibility is a dust lane preventing light from BN from reaching the 
reflection nebulae to the south.
Another dust lane crosses in front of BN at a position angle of $\sim 300^\circ$,
almost in line with BN's polarization position angle of $114^\circ$.
 
The southern region includes the bright reflection nebulae IRc3, IRc4, and IRc5,
and the controversial sources IRc2 and IRc7.
IRc2 is mostly obscured by the OMC-1 hot core at 2 \micron, but 
components IRc2-B and IRc2-D are both visible in our images.
This is the shortest wavelength at which either has been detected. 
Both sources are substantially more polarized than their immediate surroundings; 
however, their polarization position angles are similar to the polarization position angles
of the surrounding nebulosity, the polarization of which is due to dichroic absorption.
We conclude that the position angles of IRc2-B and IRc2-D 
are consistent with either dichroic absorption by foreground aligned dust grains 
or with scattered light with Star n being the illuminating source. 
{\it Neither IRc2-B nor IRc2-D is a dust cloud illuminated by a star near radio source I.}
Considering the MIR luminosities of all the sources in the region,
we suggest that both are deeply embedded, self-luminous objects 
with optically thick envelopes obscuring any central stars.
IRc7 is probably also a self-luminous object with an optically thick envelope 
scattering and polarizing the light from its central star/YSO. 

The only dust clouds whose polarization vectors are consistent with illumination by 
a star near radio source I are IRc4 and IRc5,
although this identification is not statistically significant.
These reflection nebulae could also be illuminated by components of IRc2, as is IRc3. 
We conclude that there is no strong evidence at 2 \micron\ for the object that excites 
radio source I illuminating any other object in our field of view.

In fact, there is no evidence for any illuminating source dominating the region south of BN.
The lack of evidence for such a dominating source suggests that 
no single object currently powers the strong outflows seen in CO and H$_2$.
Evidence of a dominating source, if such existed, 
would contradict theories like that of Bally \& Zinnecker (2005) 
that the outflows originated in an explosion some 500 -- 1000 yr ago.

The optically obscured stars are strongly polarized, 
indicating that they are deeply embedded 
in the cloud of dust grains aligned by the OMC-1 magnetic field.
We use the polarization position angles of the stars to investigate the OMC-1 magnetic field,
assuming that the angles are indeed parallel to the magnetic field, 
as is generally supposed.
As such, we infer that 
the magnetic field in OMC-1 changes direction from $\sim 120^\circ$ in the region north 
of IRc2 to $\sim 140^\circ$ in the region south and east of IRc2.
The objects exhibiting these position angles may be relatively foreground of the hot core --- 
south of the hot core there appears to be an additional cold cloud,
producing the polarization seen in the mm by Rao et al.\ (1998) and 
in our 2~\micron\ observations of the deeply embedded Star 25 at position angle $\sim 2^\circ$.
From the observed polarization position angles we infer that the magnetic field in this cloud 
must have a north-south orientation 
(or east-west if the grains have recently been aligned by an outflow).
Either orientation is greatly different from the overall magnetic field orientation 
foreground to the cloud.
Another location with a magnetic field orientation significantly different from the overall field
is IRc3, where again the magnetic field may be in either a north-south direction or 
an east-west direction, but $\sim 90^\circ$ different from that of the region in front of Star 25.
These variations in magnetic field direction may be connected to the strong outflows
seen in OMC-1 and centered near radio source I, IRc2, and Star n.

\acknowledgments

We thank D. C. Hines for helpful discussions of NICMOS polarimetry and for providing 
coefficients for post-NCS polarizer filter transmissions in advance of publication.
We thank J. L. Dotson and M. R. Haas for their careful reading of the manuscript
and the referee for comments which improved the quality of the manuscript.
Support for Program 9752 was provided by NASA through a 
grant from the Space Telescope Science Institute, which is 
operated by the Association of Universities for Research in 
Astronomy, Inc., under NASA contract NAS5-26555.
J.P.S acknowledges support from NASA/Ames Research Center Research Interchange Grant NCC2-1367
to SETI Institute.

{\it Facility:} \facility{HST (NICMOS)}.

\clearpage

\begin{figure}
\epsscale{1.0}
\plottwo{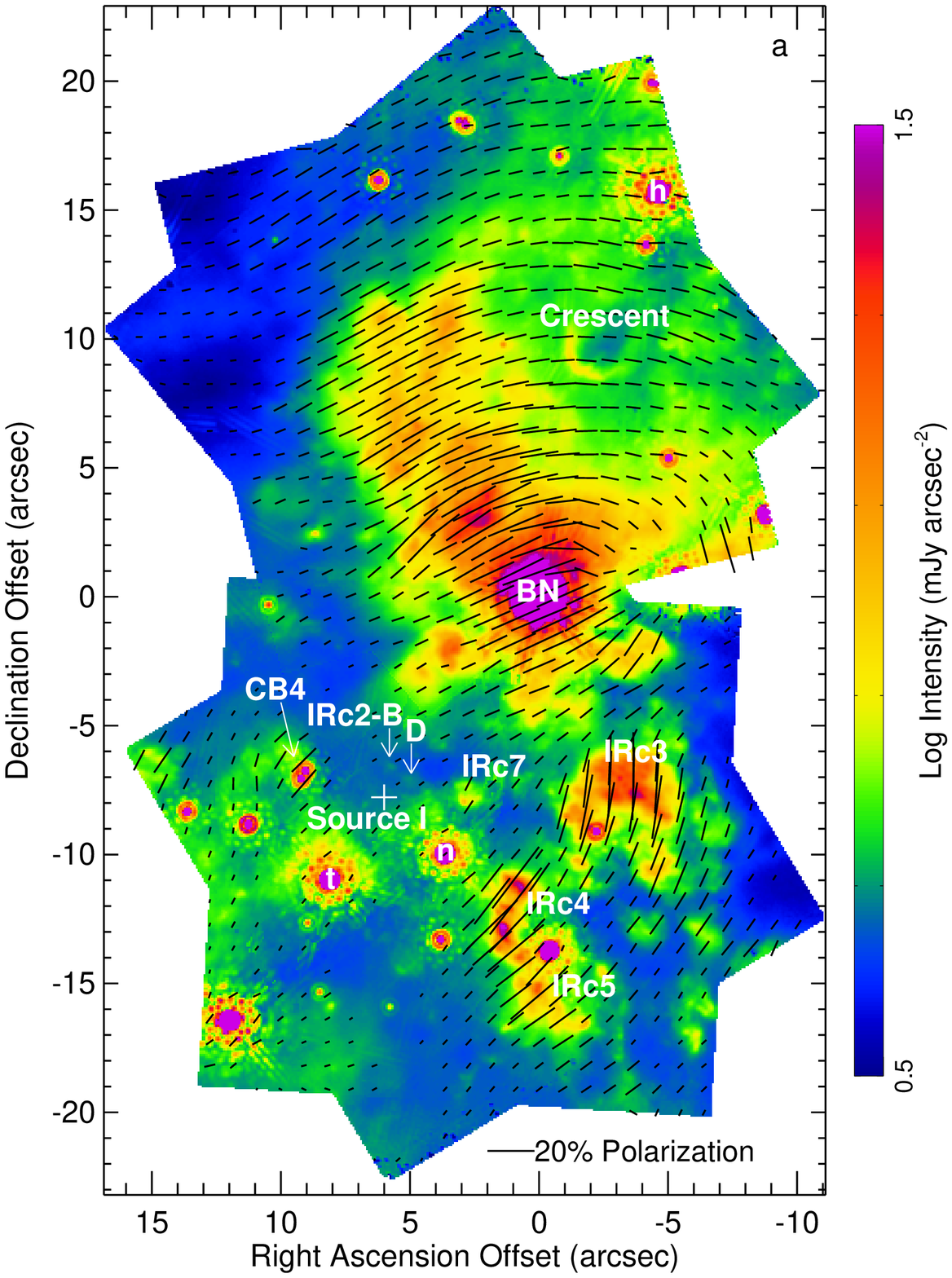}{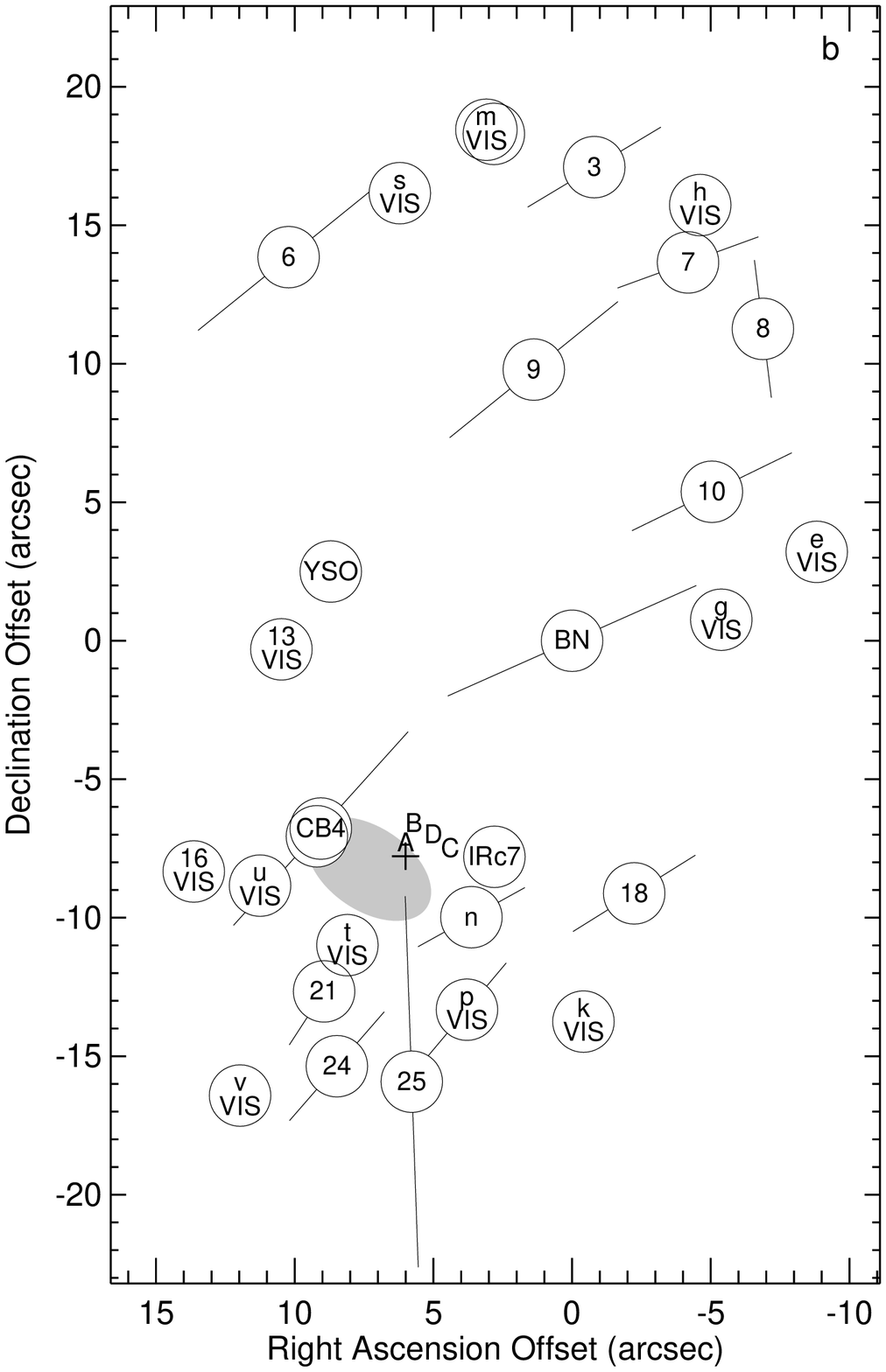}
\caption{a. The log of the Stokes intensity (total flux) for all four visits 
with polarization vectors superposed.
The scale is mJy arcsec$^{-2}$ and the plot has the brightest stars saturated.
[{\it See the electronic edition of the Journal for a color version of this figure.}]
\label{fig1}
b. Finding chart for stars and other sources of interest.
Open circles with letters are stars from Lonsdale et al.\ (1982).
Open circles with numbers are stars from Table 2 (this paper).
Stars marked ``VIS'' have been detected at visible wavelengths (Hillenbrand 1997)
and are probably foreground to OMC-1.
The lines bisecting certain stars indicate the polarization position angles for those stars 
with reliable values of $P$; the length is proportional to $P$ plus a constant.
The letters A, B, C, and D mark the locations of IRc2-A, -B, -C, and -D.
Radio source I is marked by a cross (Menten \& Reid 1995).
The gray oval marks the location of the hot core (Beuther et al.\ 2004).
}
\end{figure}

\begin{figure}
\epsscale{1.0}
\plotone{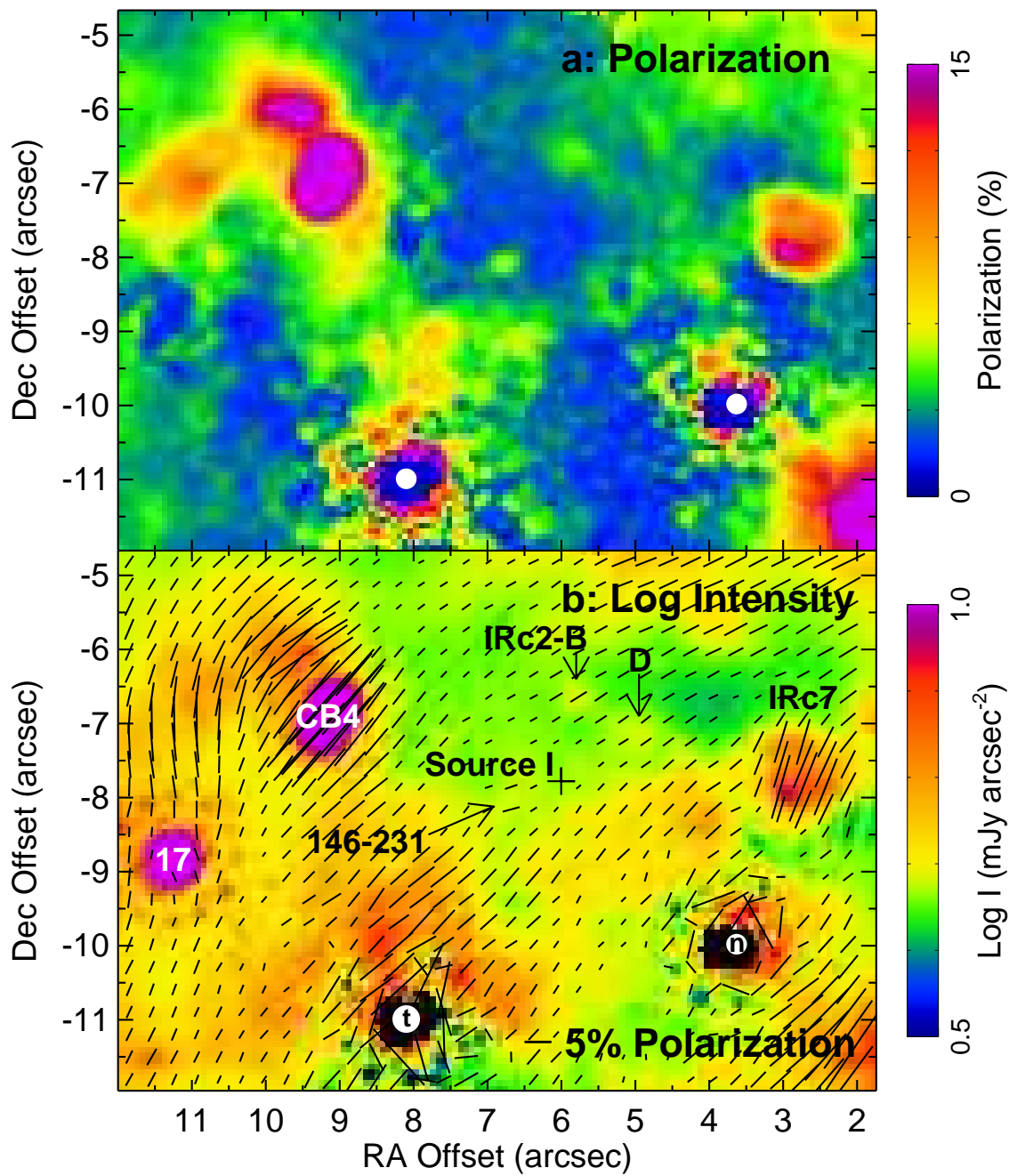}
\caption{The region including IRc2, IRc7, Stars n, t, CB4, and 17, and candidate YSO 146-231.
a. Percentage polarization. b. Log intensity.
The PSFs for Stars n and t have been subtracted and the region close to the stars blacked out
(the large polarization vectors surrounding n and t are artifacts of the 
image alignment and subtraction process).
The location of radio source I, which has been suggested to be the powering source of the hot core 
and the IRc sources, is labeled as a black cross.
[{\it See the electronic edition of the Journal for a color version of this figure.}]
\label{fig2}
}
\end{figure}

\begin{figure}
\epsscale{.80}
\plotone{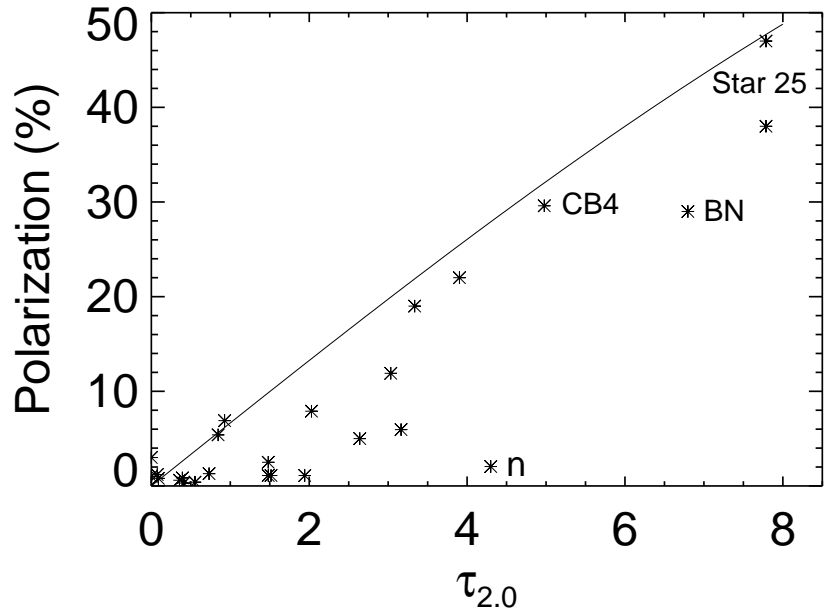}
\caption{A plot of observed polarization versus estimated $\tau_{2.0}$ (see text).
The solid line is the estimated maximum polarization $P = {\rm tanh} \tau_P$ 
for $\eta = 0.875$ from Jones (1989). 
}
\end{figure}

\begin{figure}
\epsscale{0.70}
\plotone{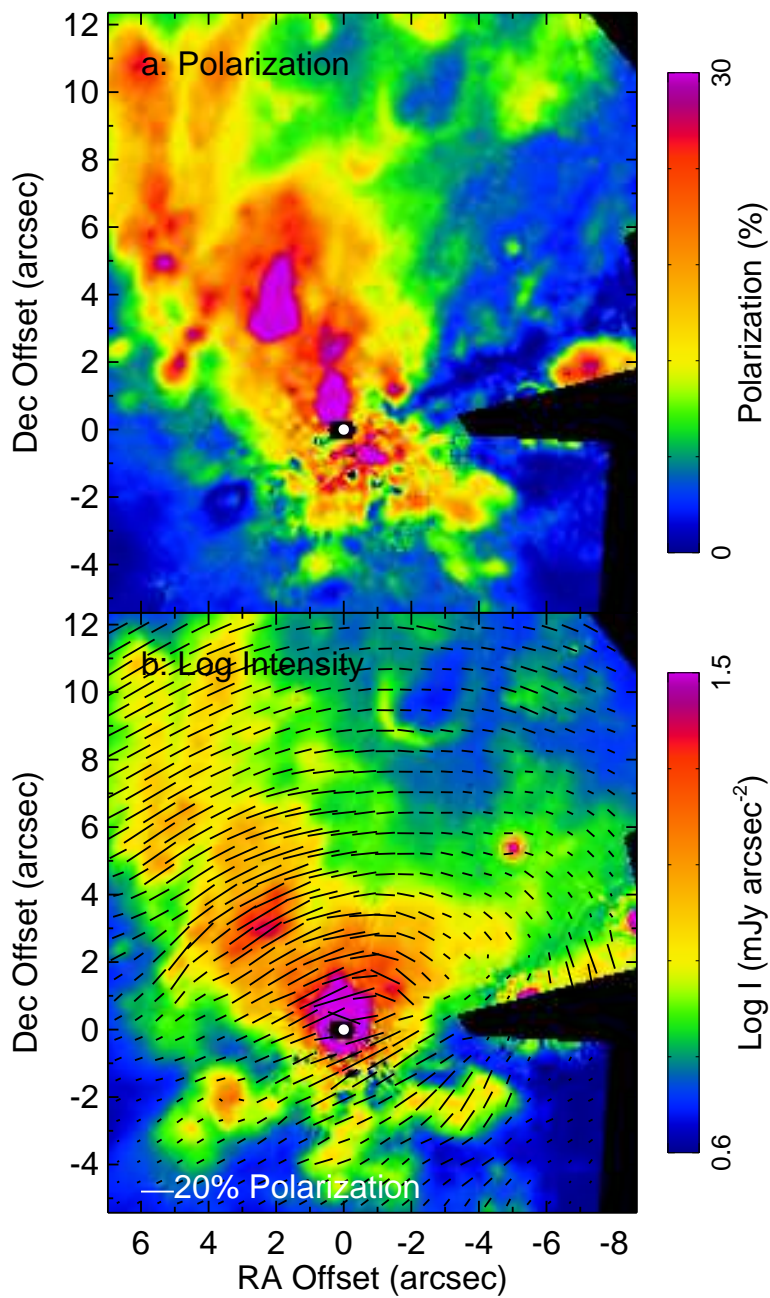}
\caption{The nebulosity surrounding the BN object. The PSF for BN, calculated using Tiny Tim, 
was subtracted. a. Percentage polarization. b. Log intensity.
Most of the diffuse material appears to be illuminated by BN. The bright, relatively unpolarized
source (143-255) at $+3.4''$, $-2.0''$ southeast of BN emits strongly in H$_2$ 
(Stolovy et al.\ 1998; Colgan et al. 2006) 
and is probably foreground to BN.
[{\it See the electronic edition of the Journal for a color version of this figure.}]
\label{fig4}
}
\end{figure}

\begin{figure}
\epsscale{1.0}
\plotone{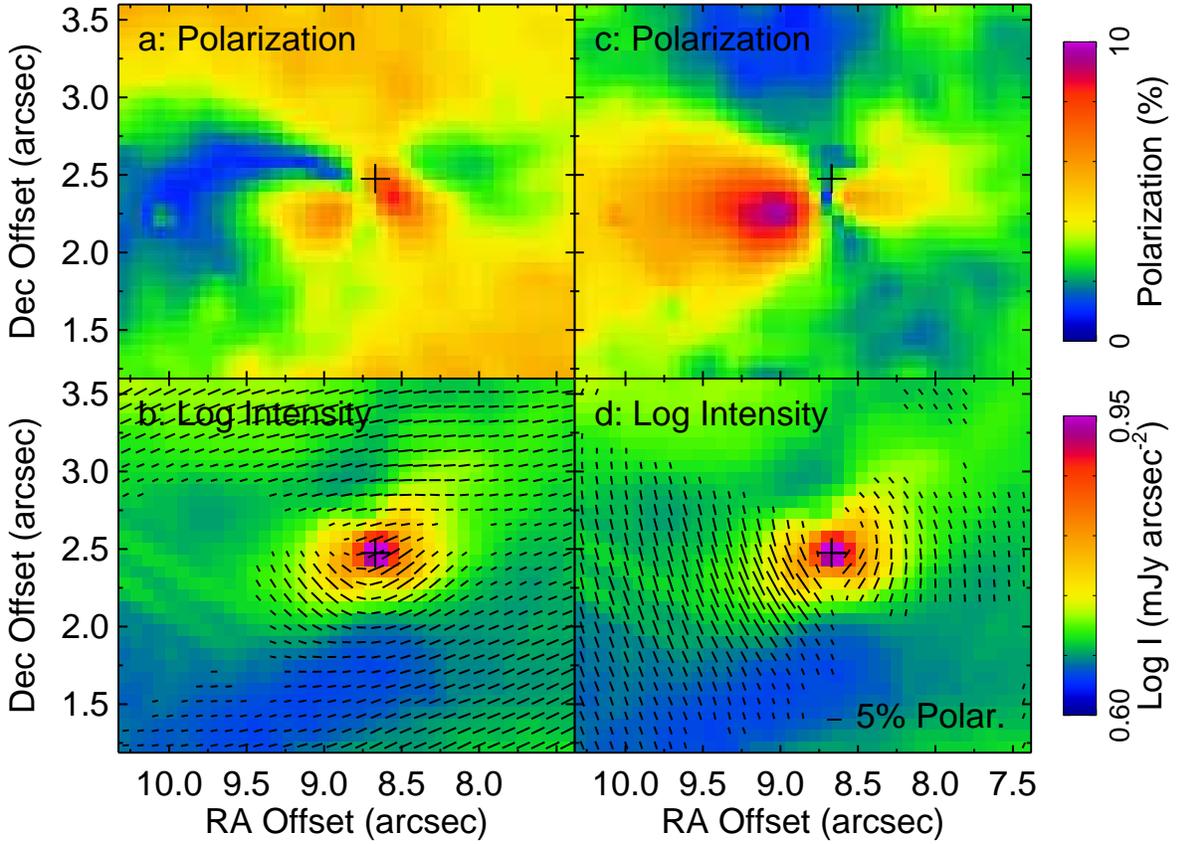}
\caption{A YSO, 147-220, with an optically thick disk and bipolar scattering cones.
A cross marks the location of the peak intensity.
a. Percentage polarization. 
b. Log intensity from Fig.\ 1a. 
c. Percentage polarization after subtraction of dichroic screen (see text).
d. The revised polarization vectors plotted on the original log I.
[{\it See the electronic edition of the Journal for a color version of this figure.}]
\label{fig5}
}
\end{figure}

\begin{figure}
\epsscale{1.0}
\plotone{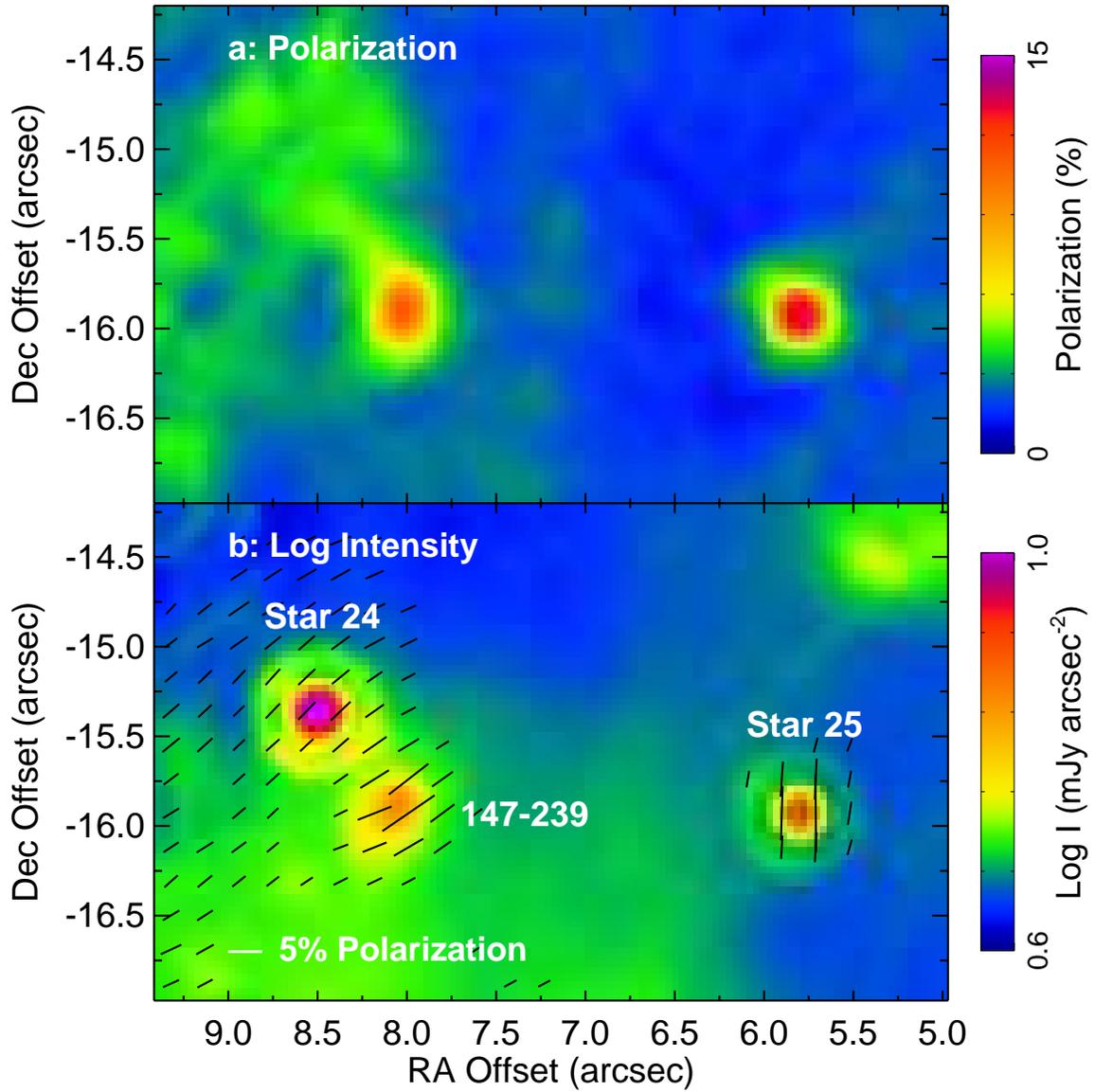}
\caption{Stars 24 and 25 plus the candidate YSO, 147-239, at $8.1''$E, $15.9''$S 
(Stolovy et al.\ 1998).
[{\it See the electronic edition of the Journal for a color version of this figure.}]
\label{fig6}
}
\end{figure}

\begin{figure}
\epsscale{1.0}
\plotone{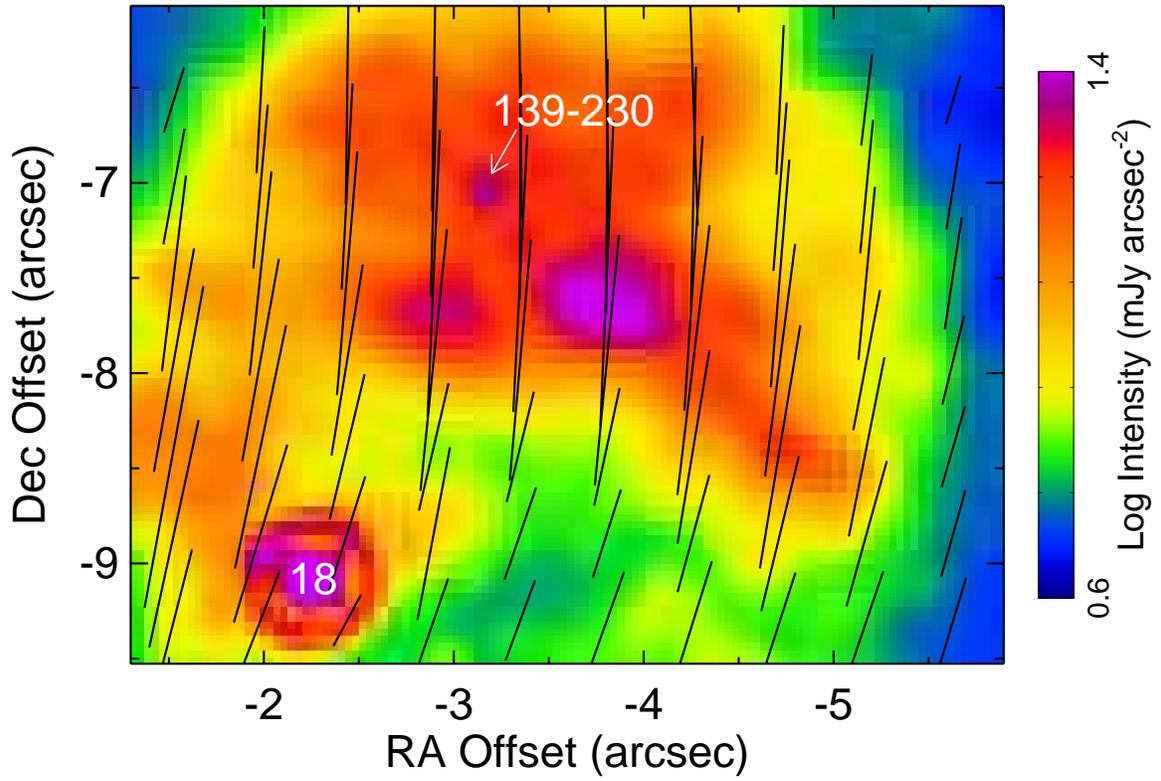}
\caption{IRc3 with candidate star, 139-230.
[{\it See the electronic edition of the Journal for a color version of this figure.}]
\label{fig7}
}
\end{figure}

\begin{figure}
\epsscale{1.0}
\plotone{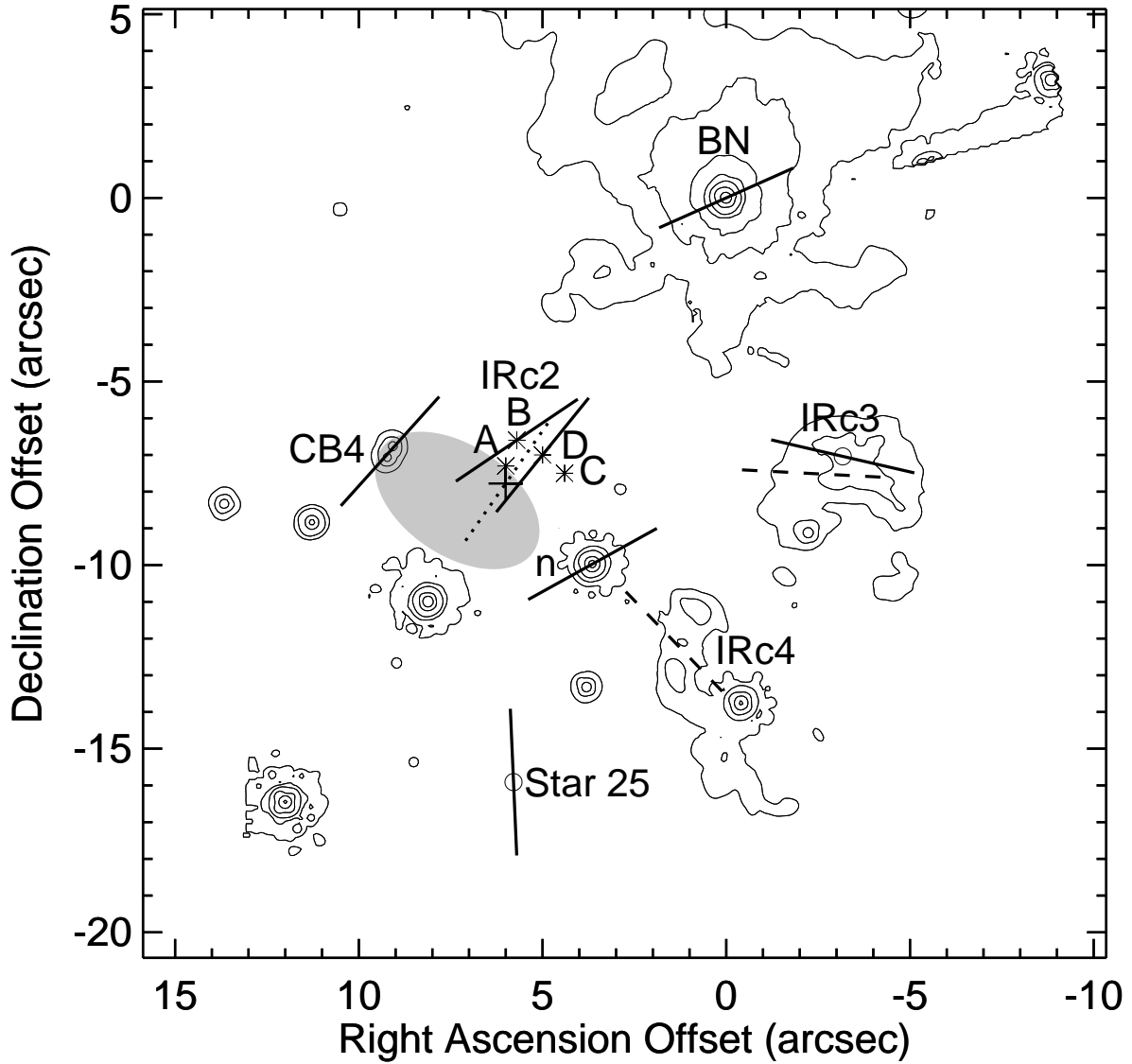}
\caption{Polarization position angles of the sources with the largest optical depths 
plotted on contours of the 2 \micron\ intensity.
The sources are BN, the candidate star in IRc3 (139-230, Table~3), 
Stars n, 25, and CB4 (Table 2), IRc2-B and IRc2-D (Table 3), 
the IRc3 and IRc4 absorptive polarization measured at 10 \micron\ 
(Aitken et al.\ 1997, dashed lines),
and the SiO masers near radio source I (Plambeck et al.\ 2003, dotted line).
The solid lines are the polarization position angles from this paper.
The location of radio source I is marked with a cross 
and the locations of the IRc2 components are plotted as asterisks.
}
\end{figure}

\clearpage


\begin{deluxetable}{ccccc} 
\tablewidth{0pt} 
\tablecaption{Journal of the Observations}
\tablehead{ 
\colhead{Visit} & \colhead{Date} & \colhead{Center RA} & \colhead{Center Dec} & \colhead{Position Angle} 
}
\startdata 
1 & 2004 Jan 12 & 5\ 35\ 14.20 & -5\ 22\ 32.0 & 176.57 \\
2 & 2004 Aug 24 & 5\ 35\ 14.22 & -5\ 22\ 31.8 & 31.16 \\
3 & 2004 Jan 20 & 5\ 35\ 14.20 & -5\ 22\ 13.0 & -165.43  \\
4 & 2004 Feb 6 & 5\ 35\ 14.20 & -5\ 22\ 13.0 & -140.43 \\
\enddata 

\end{deluxetable} 




\begin{deluxetable}{lccccccccccccc} 
\tabletypesize{\scriptsize}
\rotate
\tablewidth{0pt} 
\tablenum{2}
\tablecaption{Stellar Polarization and Photometry Measurements}
\tablehead{ 
\colhead{Star} & \colhead{LBLS\tablenotemark{a}} & \colhead{MLLA\tablenotemark{b}} & 
\colhead{HC00\tablenotemark{c}} & \colhead{H97\tablenotemark{d}} & \colhead{COUP\tablenotemark{e}} & 
\colhead{RA Off.} & \colhead{Dec Off.} & \colhead{RA} & \colhead{Dec} & 
\colhead{P (\%)} & \colhead{$\theta$} & \colhead{Mag$_{2.0 \mu m}$} & \colhead{Mag$_{2.15 \mu m}$} 
}
\startdata 
1 & m-E & 00709A & 537 & 436b\tablenotemark{f} & 620 & 3.09 & 18.45 & 5\ 35\ 14.32 & -5\ 22\ 4.31 & $1$ & \nodata & 13.28 & 12.89 \\
2 & m-W & 00709B & 537 & 436a\tablenotemark{f} & 620 &  2.82 & 18.30 & 5\ 35\ 14.30 & -5\ 22\ 4.46 & $1$ & \nodata & 13.24 & 12.48 \\
3 & \nodata & 00707 & 704 & \nodata & \nodata & -0.80 & 17.10 & 5\ 35\ 14.06 & -5\ 22\ 5.66 & $8$ & $121 \pm 1$ & 13.51 & 13.31 \\
4 & s & 00704 & 530 & 442 & 638 & 6.22 & 16.16 & 5\ 35\ 14.52 & -5\ 22\ 6.60 & $1$ & \nodata & 11.93 & 12.04 \\
5 & h & 00703 & 703 & 423 & 579 & -4.62 & 15.73 & 5\ 35\ 13.80 & -5\ 22\ 7.03 & $1$ & \nodata & 9.02 & 8.83 \\
6 & \nodata & \nodata & \nodata & \nodata & \nodata & 10.23 & 13.85 & 5\ 35\ 14.79 & -5\ 22\ 8.91 & $22$ & $129 \pm 2$ & 16.72 & 15.78 \\
7 & \nodata & 00693 & 525 & \nodata & 580  & -4.18 & 13.66 & 5\ 35\ 13.83 & -5\ 22\ 9.10 & $7$ & $110 \pm 1$ & 12.23 & 11.88 \\
8 & \nodata & 00686 & \nodata & \nodata & \nodata & -6.88 & 11.26 & 5\ 35\ 13.65 & -5\ 22\ 11.50 & $5$ & $7 \pm 9$ & 16.72 & 16.02 \\
9 & \nodata & \nodata & \nodata & \nodata & \nodata & 1.39 & 9.78 & 5\ 35\ 14.20 & -5\ 22\ 12.98 & $19$ & $129 \pm 2$ & 15.34 & 15.20 \\
10 & \nodata & 00665 & 499 & \nodata & 572 & -5.05 & 5.38 & 5\ 35\ 13.78 & -5\ 22\ 17.38 & $12$ & $116 \pm 1$ & 12.53 & 11.03 \\
11 & e & 00659 & 495 & 411 & 551 & -8.83 & 3.21 & 5\ 35\ 13.52 & -5\ 22\ 19.55 & $<1$ & \nodata & 10.39 & 10.36 \\
12 & BN & 00642 & 705 & \nodata & 599 & 0 & 0 & 5\ 35\ 14.11 & -5\ 22\ 22.76 & $29$ & $114 \pm 1$ & 6.92 & 5.41 \\
13 & \nodata & 00639 & 773 & 9074\tablenotemark{f} & 661 & 10.48 & -0.31 & 5\ 35\ 14.81 & -5\ 22\ 23.07 & $1$ & \nodata & 14.48 & 14.07 \\
14=CB4N\tablenotemark{g} & u\tablenotemark{h} & 00614B & 465 & \nodata & 655 & 9.07 & -6.77 & 5\ 35\ 14.71 & -5\ 22\ 29.53 & $27$ & $138 \pm 1$ & 12.14 & 11.25 \\
15=CB4S\tablenotemark{g} & u\tablenotemark{h} & 00614A & 464 & \nodata & 655 & 9.21 & -7.06 & 5\ 35\ 14.72 & -5\ 22\ 29.82 & $32$ & $138 \pm 1$ & 12.24 & 11.35 \\
16 & \nodata & 00606 & 456 & 9086 & \nodata & 13.66 & -8.33 & 5\ 35\ 15.02 & -5\ 22\ 31.09 & $<1$ & \nodata & 12.09 & 11.72 \\
17 & u & 00603 & 453 & 452 & 663 & 11.26 & -8.83 & 5\ 35\ 14.86 & -5\ 22\ 31.59 & $<1$ & \nodata & 11.26 & 11.06 \\
18 & \nodata & 00602 & 451 & \nodata & 590 & -2.24 & -9.11 & 5\ 35\ 13.96 & -5\ 22\ 31.87 & $6$ & $122 \pm 2$ & 12.01 & 11.57 \\
19 & n & 00598 & 448 & \nodata & 621 & 3.64 & -9.97 & 5\ 35\ 14.35 & -5\ 22\ 32.73 & $2$ & $119 \pm 1$ & 9.72 & 8.78 \\
20 & t\tablenotemark{i} & 00595 & 443 & 448 & 648 & 8.11 & -10.99 & 5\ 35\ 14.65 & -5\ 22\ 33.75 & $1$ & \nodata & 9.25 &9.07 \\
21 & \nodata & 00587 & 755\tablenotemark{j} & \nodata & \nodata & 8.96 & -12.66 & 5\ 35\ 14.71 & -5\ 22\ 35.42 & $3$ & $146 \pm 10$ & 14.96 & 14.40 \\
22 & p & 00583 & 439 & 9063 & 622 & 3.79 & -13.31 & 5\ 35\ 14.36 & -5\ 22\ 36.07 & $2$ & $140 \pm 4$ & 12.01 & 11.66 \\
23 & k & 00581 & 438 & 432 & 600 & -0.41 & -13.74 & 5\ 35\ 14.08 & -5\ 22\ 36.50 & $<1$ & \nodata & 10.02 & 9.80 \\
24 & \nodata & 00572B\tablenotemark{k} & 757 & \nodata & \nodata & 8.50 & -15.35 & 5\ 35\ 14.68 & -5\ 22\ 38.11 & $6$ & $140 \pm 2$ & 14.99 & 14.39 \\
25 & \nodata & 00570\tablenotemark{l} & \nodata & \nodata & 639 & 5.79 & -15.91 & 5\ 35\ 14.49 & -5\ 22\ 38.67 & 47V1 & $2 \pm 1$ & 16.13V1 & 17.48 \\
 &&&&&&&&&& 38V2 & $3 \pm 1$ & 15.84V2 & \\
26 & v & 00568 & 431 & 454 & 670 & 11.97 & -16.42 & 5\ 35\ 14.90 & -5\ 22\ 39.18 & $1$ & \nodata & 8.87 & 8.54 \\
\enddata 


\tablenotetext{a}{~LBLS: Lonsdale et al. (1982).}

\tablenotetext{b}{~MLLA: Muench et al. (2002), surveys taken 1997 -- 2000.}

\tablenotetext{c}{~HC00: Hillenbrand \& Carpenter (2000), survey taken 1999 Feb.}

\tablenotetext{d}{~H97: Hillenbrand (1997), survey in V and I$_C$ bands. 
H97 numbers from 1 -- 1053 are taken from Jones \& Walker (1988); 3000 -- 4999 were measured in Jan 1993; 
5000 -- 5999 were measured in Feb. 1995; 6000+ were measured in Feb. 1996.}

\tablenotetext{e}{~{\it Chandra} Orion Ultradeep Project (COUP): Grosso et al. (2005).}

\tablenotetext{f}{~Star is not a member of the Orion Nebula Cluster (ONC, H97).}

\tablenotetext{g}{~These stars have more uncertain magnitudes and polarization ($\pm 3$\%) than the others because their PSFs overlap; 
however, the polarization and brightness differences are real.}

\tablenotetext{h}{~LBLS used a $3.5''$ beam to scan the OMC-1 region. Stars 14, 15, and 17 are 
separated by only $2.8''$, so it is perhaps not that surprising that LBLS did not see
two separate stars at the location they call ``u''. They give magnitudes of H=11.7, K=10.6, L=7.0,
and V=17 for LBLS u. Surely the V magnitude refers to star 17 only but their NIR magnitudes 
are for the sum of the three stars.
The coordinate differences of u from BN are $10.3''$ and $8.7''$. 
This location is halfway between CB4 and star 17 in RA but much closer to star 17 in Dec.
We use the designation ``u'' now to refer only to star 17.
}

\tablenotetext{i}{~Menten \& Reid (1995) detect a weak radio source at the location of LBLS-t.}

\tablenotetext{j}{~This star is much fainter now (January, 2004) than it was 
when it was observed by HC00, who measured K = 12.436 and H = 12.299 in February, 1999.
It was also fainter when the F215N data were taken on 13 April, 1997.}

\tablenotetext{k}{~MLLA 00572A is not a NICMOS point source --- it is the candidate YSO 147-239 in Table 3.}

\tablenotetext{l}{~MLLA flag their K$_s$ photometry of MLLA 00570 as ``likely corrupted''.}

\end{deluxetable} 




\begin{deluxetable}{lcccc} 
\tablewidth{0pt} 
\tablenum{3}
\tablecaption{Extended Source Polarization Measurements}
\tablehead{ 
\colhead{Source} & \colhead{RA Offset\tablenotemark{a}} & \colhead{Dec Offset\tablenotemark{a}} & \colhead{P (\%)} & \colhead{$\theta$}  
}
\startdata 
146-231 & 6.8 & $-8.1$ &  $75 \pm 25$ & $88 \pm 8$ \\
147-239  & 8.1 & $-15.9$ & $37 \pm 2$ & $121 \pm 4$ \\
IRc2-B & 5.7 & $-6.6$ & $35 \pm 9$ & $124 \pm 2$  \\
IRc2-D & 5.0 & $-7.0$ & $10 \pm 3$ & $139 \pm 6$ \\
IRc3NE & $-2.5$ & $-6.5$ & $26 \pm 2$ & $179 \pm 1$ \\
IRc3E  & $-1.5$ & $-8.5$ & $35 \pm 1$ & $169 \pm 2$ \\
IRc3S  & $-3.5$ & $-7.5$ & $26 \pm 2$ & $179 \pm 1$ \\
IRc3SW & $-4.5$ & $-10.2$ & $24 \pm 1$ & $160 \pm 5$ \\
IRc4N  & 0.5 & $-11.2$ & $43 \pm 4$ & $143 \pm 3$ \\
IRc4E  & 1.3 & $-13.0$ & $57 \pm 3$ & $134 \pm 3$ \\
IRc5  & 0.0 & $-15.0$ & $40 \pm 2$ & $133 \pm 3$ \\
IRc5SW & $-1.5$ & $-16.5$ & $5 \pm 1$ & $129 \pm 3$ \\
IRc7 & 2.8 & $-7.7$ &  $31 \pm 4$ & $161 \pm 3$ \\
139-230 & $-3.16$ & $-7.04$ & $20 \pm 6$ & $77 \pm 6$ \\
\enddata 


\tablenotetext{a}{Offset positions from BN in arcsec.}

\end{deluxetable} 




\begin{deluxetable}{lcccc} 
\tablewidth{0pt} 
\tablenum{4}
\tablecaption{Known and Suggested Properties of Objects in OMC-1}
\tablehead{ 
\colhead{Object} & \colhead{Illuminating Source} & \colhead{Object Type} & \colhead{P\tablenotemark{a} (\%)} & \colhead{$\theta$\tablenotemark{a}}  
}
\startdata 
BN & Self & Star & $29$ & $114$  \\
Star n & Self & Star & $2$ & $119$ \\
Source I & ? & Radio Source & \nodata & \nodata \\
IRc2-B & Self? & YSO? & $35$ & $124$  \\
IRc2-D & Self? & YSO? & $10$ & $139$ \\
IRc3 & IRC2-A & RN & $26$ & $178$ \\
IRc4  & I or IRc2 & RN & $33$ & $139$ \\
IRc5  & I or IRc2 & RN & $17$ & $127$ \\
IRc7 & Self? & YSO? &  $31$ & $161$ \\
CB4 & Self & Binary Stars & $27, 32$ & $138, 138$  \\
Star 25 & Self & Star & $47, 38$ & $2, 3$  \\
147-220 & Self & YSO &  See Fig. 5 & See Fig. 5 \\
146-231 & Self? & YSO? &  $75$ & $88$ \\
147-239  & Self? & YSO? & $37$ & $121$ \\
139-230 & Self? & YSO? & $20$ & $77$ \\
\enddata 


\tablenotetext{a}{At 2.0 \micron.}

\end{deluxetable} 


\end{document}